\documentclass[jcp,twocolumn,aip,reprint,superscriptaddress,amssymb,samsmath,showpacs,floatfix]{revtex4}
\usepackage{graphicx}
\usepackage{multirow}
\usepackage{amsmath}
\usepackage{color}

\usepackage{subfigure}
\usepackage{epstopdf}
\def\UOM#1{\,\hbox{#1}}

\begin{document}

\title{%Computer Simulations of End-to-End Stacking of Pair DNA Duplexes
Self-assembly of short DNA duplexes: from a coarse-grained model to experiments through a theoretical link}
\author{Cristiano~De Michele\,$^{1,}$\footnote{To whom correspondence should be addressed.
Tel: +390649913524; Fax: +39064463158; Email: cristiano.demichele@roma1.infn.it},
Lorenzo Rovigatti\,$^{1}$,
Tommaso Bellini\,$^{2}$ 
and Francesco Sciortino\,$^{3}$%
}
\address{
$^{1}$ Dipartimento di Fisica, ``{\em Sapienza}'' Universit\`a di Roma, P.le A. Moro 2, 00185 Roma, Italy
$^{2}$ Dipartimento di Chimica, Biochimica e Biotecnologie per la Medicina, Universit\`a di Milano, I-20122 Milano, Italy
and 
$^{3}$ Dipartimento di Fisica  and  CNR-ISC, ``{\em Sapienza}'' Universit\`a di Roma, P.le A. Moro 2, 00185 Roma, Italy
}
\date{\today}

\begin{abstract}
Short blunt-ended DNA duplexes comprising 6 to 20 base pairs self-assemble into polydisperse semi-flexible chains
due to hydrophobic stacking interactions between terminal base pairs. Above a critical concentration, which depends
on temperature and duplex length, such chains order into liquid crystal phases. Here, we investigate the self-assembly of such double-helical duplexes with a combined  numerical and theoretical approach. We simulate the bulk system  employing the coarse-grained  DNA model  recently proposed by Ouldridge {\it et al.} [ J. Chem. Phys.  {\bf 134},  08501 (2011) ].  Then we evaluate the input quantities for the theoretical framework directly from the DNA model.  The resulting parameter-free theoretical predictions provide an accurate description of the simulation results in the isotropic phase.  In addition, the theoretical isotropic-nematic phase boundaries are in line with experimental findings, providing a route to estimate  the stacking free energy.
\end{abstract}
%\pacs{64.70.mf,61.30.Cz,64.75.Yz,87.15.A-,82.35.Pq,87.14.gk}

\maketitle
\section{Introduction}
%Self-assembly, that is the spontaneous formation of reversible aggregates, is ubiquitous in nature and is relevant in several fields, including material science, soft matter and biophysics~\cite{HamleyBook,Glotz_04,WhitesidesPNAS02}. Indeed, self-assembly may involve the structuring of simple molecules, of macromolecules or of colloidal particles, and thus it takes place at length scales ranging from angstrom to microns. Controlling the self-assembly by tuning the properties and interactions of the aggregating units is an important tool for the design of new structures~\cite{kumachevaNatMat07,doye,Cho_05,Workum_06,dnastarr,MIRKIN_96}. 

Self-assembly is the spontaneous formation through free energy minimization of reversible aggregates of basic building blocks.  The  size
of the aggregating units, e.g. simple molecules, macromolecules or colloidal particles, can vary  from a few angstr\"oms to microns, 
thus making self-assembly ubiquitous in nature and of interest  in several fields, including material science, soft matter and  biophysics~\cite{HamleyBook,Glotz_04,WhitesidesPNAS02,Cho_05,dnastarr}. Through self-assembly it is possible to design new materials whose physical properties are controlled by tuning the interactions of the individual building blocks~\cite{kumachevaNatMat07,doye,MIRKIN_96,Workum_06}.

A  relevant self-assembly process   is the formation of  filamentous aggregates 
(i.e. linear chains) induced by the anisotropy of attractive interactions. 
Examples  are provided by micellar systems~\cite{Khan96,CatesLangmuir94,KnutzSM08},
formation of fibers and fibrils~\cite{MezzengaLangmuir2010,LeePRE09,CiferriLC07,AggeliJACS03},  solutions of long duplex B-form DNA composed of $10^2$ to $10^6$ base pairs ~\cite{RobinsonTetra61,LivolantNature89,MerchantBJ97,FerrariniJCP05}, filamentous viruses~\cite{FerrariniPRL06,DogicSM09,FredenPRL03,TomarJACS07,MinskyNatCell02},
chromonic liquid crystals~\cite{LydonJMC10} as well as inorganic nanoparticles~\cite{eugenia-science}.

If linear aggregates possess sufficient rigidity, the system may exhibit liquid
crystal (LC) phases (e.g. nematic or columnar) above a critical concentration.
In the present study we focus on the self-assembly of short (i.e. 6 to 20 base pairs) DNA duplexes 
(DNADs)~\cite{BelliniScience07,BelliniJPCM08,BelliniPNAS2010} 
in which coaxial stacking interactions between the blunt ends of the DNADs favor their aggregation into weakly bonded chains.  Such a reversible physical polymerization is enough to promote the mutual alignment of these chains and the formation of macroscopically orientationally ordered nematic LC phases. 
At present, stacking is understood in terms of hydrophobic forces acting between the flat hydrocarbon surfaces provided by the paired nucleobases at the duplex terminals, although the debate on the physical origin 
of such interaction is still active and lively~\cite{KoolJACS00, BelliniReview2011}. In this respect, the self-assembly of  DNA duplexes provides a suitable way to access and quantify hydrophobic coaxial stacking interactions. 

In order to extract quantitative informations from DNA-DNA coaxial stacking experiments, reliable computational models and theoretical frameworks are needed. Recent theoretical approaches have focused on  the  isotropic-nematic (I-N) transition in self-assembling systems
~\cite{GlaserMC,KindtJCP04}, building on previous work on rigid and semi-flexible polymers~\cite{Vroege92,DijkstraPRE97,Semenov81,Semenov82,MulderJPCM06,DijkstraPRL11,WesselsSM03,ChenMacromol93,Odijk86}.  In a recent publication~\cite{ourMacromol} we investigated the
reversible physical polymerization and collective ordering of DNA duplexes by modeling them as super-quadrics
with quasi-cylindrical shape~\cite{mioJCP} with two reactive  sites~\cite{SMcorezzi09,corezziJPCB}  on their bases. 
Then we presented a theoretical framework, built on Wertheim~\cite{WertheimJSP1,WertheimJSP2,WertheimJSP3} and Onsager~\cite{Onsager49} theories, which is able to properly  account for the association process. 

Here, we employ this theoretical framework to study the physical properties
of a realistic coarse-grained model of DNA recently proposed by Ouldridge {\it et al.}~\cite{ouldridge_jcp}, where nucleotides are modeled as rigid bodies interacting with site-site potentials.  
 Following Ref.~\cite{ourMacromol}, we compute 
the inputs required by the theory, %of  Ref.~\cite{ourMacromol}, 
i.e. the stacking  free energy and the DNAD excluded volume, for the Ouldridge {\it et al.} model \cite{ouldridge_jcp}. Subsequently we predict the polymerization extent in the isotropic phase as well as the isotropic-nematic phase boundaries. 

To validate the theoretical predictions, we perform large-scale molecular dynamics (MD) simulations in the NVT ensemble of a bulk system comprising $9600$ nucleotides,   a study made possible by the  computational power of modern Graphical Processing Units (GPUs). 
The parameter-free theoretical predictions provide an accurate description of the simulation results 
in the isotropic phase, supporting the theoretical approach and its application in the comparison with experimental results.
 
The article is organized as follows. Section~\ref{sec:methods} provides details of the coarse-grained model of DNADs, of the MD computer simulations and
presents a summary of the theory.
Section~\ref{sec:results} describes the protocols implemented to evaluate the input parameters required by 
the theory via MC integrations for two DNADs. We also discuss some geometrical properties of the bonded dimer configurations.  We then compare the theoretical predictions  with  simulation and experimental results.  Finally, in Section~\ref{sec:conclusions} we discuss    estimates for the stacking free energy and present our conclusions.

\section{Methods}
\label{sec:methods}

\subsection{Model}
\label{sec:model}
\begin{figure*}
\centering
\includegraphics[width=18cm]{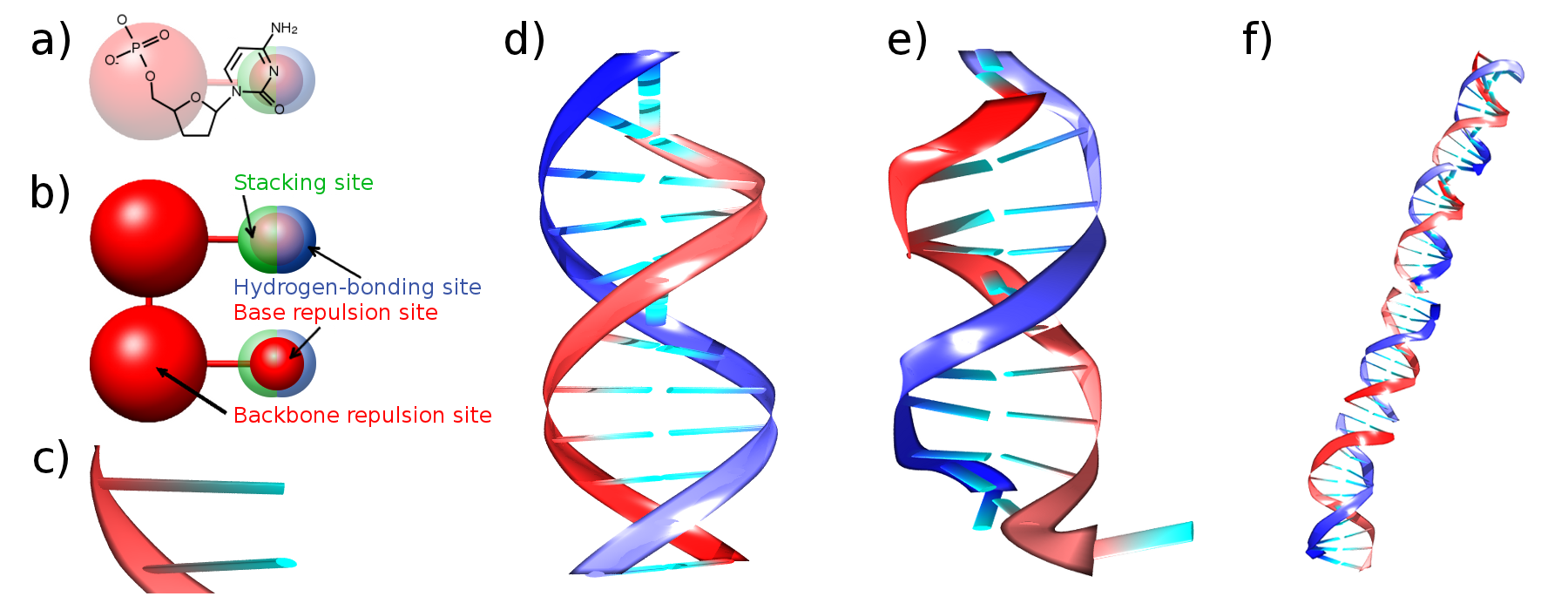}
\caption{(a) Schematic representation of the coarse graining of the model for a single nucleotide. 
(b) Model interaction sites. For the sake of clarity, stacking and hydrogen-bonding sites are highlighted 
on one nucleotide and the base repulsion site on the other. For visualization reasons, in the following 
strands will be shown as ribbons and bases as extended plates as depicted in (c).
(d) A $12$ base-pairs DNAD in the minimum energy configuration. (e) An equilibrium configuration of the same 
object at $T = 300\UOM{K}$. The nucleotides at the bottom are not bonded, the so-called fraying effect. (f) 
A chain of length $N_b=48$ extracted from a simulation at $c = 200\UOM{mg/ml}$.}
\label{fig:model}
\end{figure*}

We implement a coarse-grained model for DNA recently developed by Ouldridge \textit{et al.}~\cite{ouldridge_prl, ouldridge_jcp}. 
In such model, designed via a top-down approach, each nucleotide is described as a rigid body 
(see Figure~\ref{fig:model}). The interaction forms and parameters are chosen so as to
reproduce structural and thermodynamic properties of both single- (ssDNA) and double- (dsDNA) stranded 
molecules of DNA in B-form. All interactions between nucleotides are pairwise and, in the last 
version of the model~\cite{ouldridge_jcp}, continuous and differentiable. Such feature makes the model convenient %feasible 
foe MD simulations.

The interactions between nucleotides account for excluded volume, backbone connectivity, Watson-Crick 
hydrogen bonding, stacking, cross-stacking and coaxial-stacking.  The interaction parameters have been adjusted in order to be consistent with experimental data~\cite{ouldridge_jcp,santalucia,Holbrook}. 
%In particular lengths were initially chosen by hand to give our approximate B-DNA ge- ometry.
%<<<<<<< .mine
%In particular the stacking interaction strength and stiffness have be chosen to be consistent with the experimental thermodynamics reported for 14-base oligomers by Holbrook {\it et al.}~\cite{Holbrook}. Hydrogen-bonding and cross-stacking potentials were adjusted to give duplex and hairpin formation thermodynamics consistent with the SantaLucia parameterization of the nearest-neighbor model~\cite{santalucia}.
%=======
In particular, the stacking interaction strength and stiffness have been chosen to be consistent with the experimental results reported for 14-base oligomers by Holbrook {\it et al.}~\cite{Holbrook}. Hydrogen-bonding and cross-stacking potentials were adjusted to give duplex and hairpin formation thermodynamics consistent with the SantaLucia parameterization of the nearest-neighbor model~\cite{santalucia}.
%>>>>>>> .r343
%43 which can be viewed as an accurate empirical fit to experimental data. For compari-
%son with Ref. 43 we considered an average base pair step see Sec. III B 2as our model contains limited sequence de- pendence. 
Interaction stiffnesses were also further adjusted in order to provide correct mechanical properties of the model, such as the persistence length. The model does not have any sequence dependence apart from the Watson-Crick pairing, meaning that the strength of the interactions 
acting between nucleotides is to be considered as an average value. In addition, the model assumes  conditions of high salt molarity ($0.5 \UOM{M}$). In this model, the coaxial-stacking interaction acts between any two non-bonded nucleotides and is responsible for the duplex-duplex bonding. It has been parametrized~\cite{tesi_tom} to reproduce  experimental data which quantify the stacking interactions by observing the difference in the relative mobility of a double strand  where one of the  two strands has been nicked with respect to intact DNA~\cite{JMolBiolGST04,NuclAcResGST06} and by analyzing the melting temperatures of short duplexes adjacent to hairpins~\cite{NuclAcResGST00}.
 
To cope with  the complexity of the model and the large number of nucleotides involved in bulk simulations, we employ a modified  version of the CPU-GPU code used in a previous work~\cite{rovigatti_molphys}, and extend it to support the force-fields~\cite{allen2006expressions}. Harvesting the power of modern Graphical 
Processing Units (GPUs) results in a $30$-fold speed-up. The CPU version of the code, as well as the 
Python library written to simplify generation of initial configurations and post-processing analysis, 
will soon be released as free software~\cite{dna_code}.

\subsection{Bulk simulations}
\label{subsec:bulk_sim}

To compare numerical results with theory, we perform Brownian dynamics simulations of 
$400$ dsDNA molecules made up by $24$ nucleotides each, i.e. $400$ 
cylinder-like objects with an aspect ratio of $\approx 2$ (see Figure~\ref{fig:model}~(d)). 
The integration time step has been chosen to be $0.003$ in simulation units which corresponds, if 
rescaled with the units of length, mass and energy used in the model, to approximately $1\times 10^{-14}$ seconds. 

We study systems at three different temperatures, namely $T = 270\UOM{K}$, $285\UOM{K}$ and $300\UOM{K}$, and for different 
concentrations, ranging from $2 \UOM{mg/ml}$ to $241\UOM{mg/ml}$.  The $T=270\UOM{K}$ state point, despite being far from the
experimentally accessed $T$, is here investigated  to test the theory in a region of the phase diagram where the degree of association is significant.  To quantify  the aggregation process we define two DNADs as bonded if their pair interaction energy is negative.  Depending on temperature and concentration, we use $10^6 - 10^7$ MD steps for equilibration and $10^8 - 10^9$  MD steps for data generation on NVIDIA Tesla C2050 GPUs, equivalent to $1 - 10\;\mu s$.

\begin{figure}
\centering
\includegraphics[width=7cm]{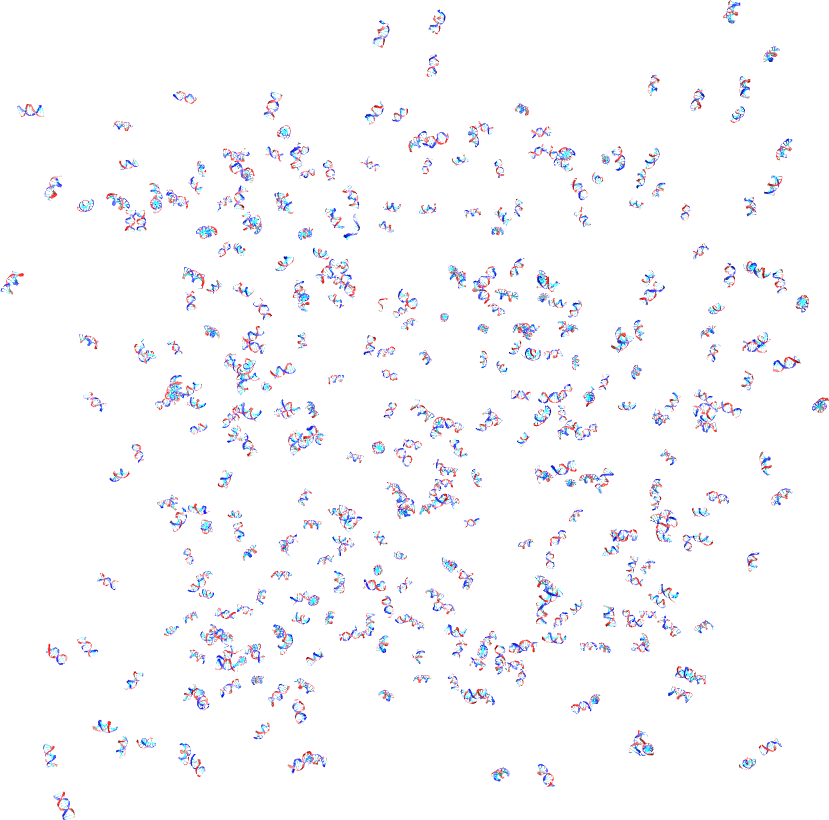}
\includegraphics[width=7cm]{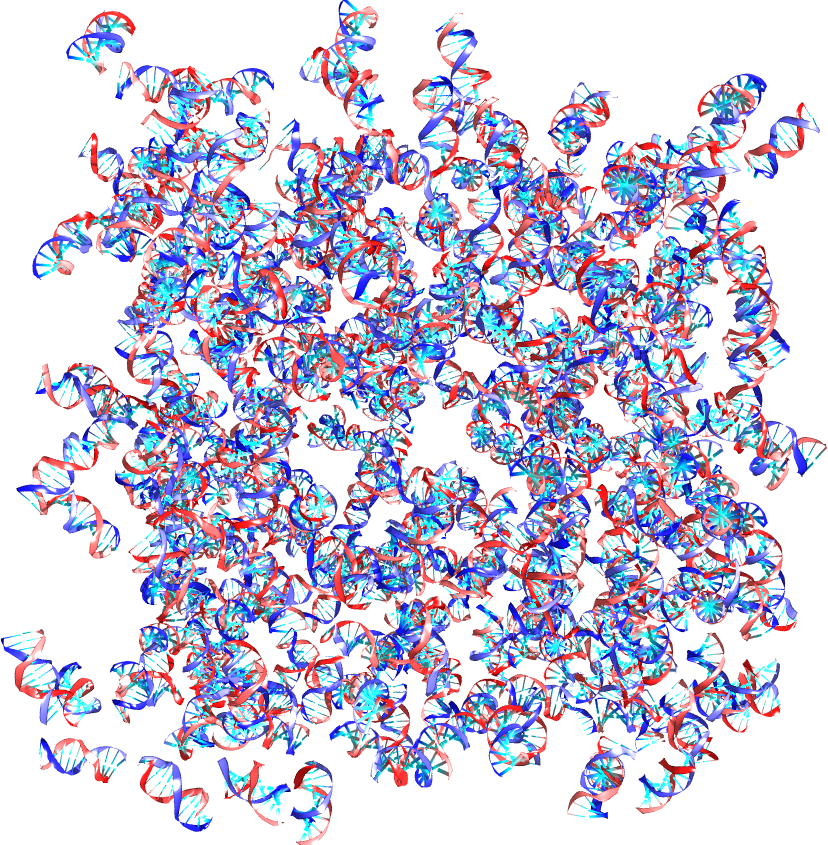}
\caption{Snapshots taken from simulations at $T = 300\UOM{K}$. At low concentrations ($c=2 \UOM{mg/ml}$, top) chain formation is negligible and the average chain length is approximately $1$. As the concentration is increased ($c=80 \UOM{mg/ml}$, bottom), DNADs start to self-assemble into chains and the average chain length increases.}
\label{fig:snapshots}
\end{figure}
\section{Theory}
\label{sec:theory}
%Following the work of van der Schoot and Cates~\cite{CatesLangmuir94,CatesEPL94} and its extension to higher volume fractions
%with the use of the Parsons-Lee approximation~\cite{Parsons79,Lee87} as suggested by Kuriabova {\it et al.}~\cite{GlaserMC},

We build on the theoretical framework previously developed to account for the linear aggregation and collective ordering of quasi-cylindrical particles~\cite{ourMacromol}. Here, we provide a discussion of how such a theory can be used to describe
the reversible chaining and ordering of oligomeric DNADs at the level of detail adopted by the present model.  According to Ref.~\cite{ourMacromol}, the free energy of  a system of equilibrium polymers can be written as
%but taking into account also the Parsons-Lee  approximation to extend the theory also to high volume 
%fractions as also done by Glaser {\it et al.}\cite{glaserMC} 
%we assume the following expression for the free energy of our system:
\begin{eqnarray}
\frac{\beta F}{V} &=& \sum_{l=1}^{\infty} \nu(l) \left \{ \ln \left [ v_d \nu(l) \right ] - 1 \right \} +\nonumber \\
&+& \frac{\eta(\phi)}{2} \sum_{l=1 \atop l'=1}^{\infty} \nu(l) \nu(l') v_{excl}(l, l')  
 \nonumber \\ 
 &-& \beta \Delta F_b \sum_{l=1}^{\infty} (l-1) \nu(l)  
 + \sum_{l=1}^{\infty} \nu(l) \sigma_o(l) 
\label{eq:freeene}
\end{eqnarray}

where $V$ is the volume of the system,   $\phi\equiv v_d \rho$  ($\rho=N/V$ is the number density of monomers) is the packing fraction, $\nu(l)$ is the number density of chains of length $l$, 
normalized such that $\sum_{l=1}^{\infty} l\, \nu(l)= \rho$, $v_d$ is the volume of a monomer,
%$f_l({\bf u}) = \nu(l) f({\bf u})$, 
$\Delta F_b$, as discussed in the subsection \ref{sec:twobody}, is a parameter which depends on the free energy associated to a single bond and
%,i.e. the stacking free energy,
 %which comprises an entropic and an energetic contribution, % $f^{id}_{i}$ is the concentration independent contribution, 
 $v_{excl}(l,l')$ is the excluded volume of two chains of length $l$ and $l'$. $\eta(\phi)$ is the Parsons-Lee factor~\cite{Parsons79}
\begin{equation}
\eta(\phi) = \frac{1}{4} \frac{4-3\phi}{(1-\phi)^2}
\label{eq:parsonlee}
\end{equation}
and $\sigma_o(l) $\cite{Odijk86}  accounts for the orientational entropy that a chain of length $l$ loses in the nematic phase (including possible contribution due to its flexibility). The explicit form for $\sigma_o(l) $ can be found in Ref.~
\cite{ourMacromol}.
 
The free energy functional (Eq.~\ref{eq:freeene})  explicitly accounts for the polydispersity inherent in the equilibrium polymerization using a discrete chain length distribution and for the entropic and energetic contributions of each single bond  through the parameter $\Delta F_b$. 

\subsubsection{Isotropic phase}
\label{sec:isophase}
\noindent In the isotropic phase $\sigma_0=0$ and
the excluded volume can be written as follows (see Appendix~\ref{sec:appendix_a}):
\begin{equation}
%v_{excl}(l,l') &=& \frac{\pi^2}{8} D^3 + \left (\frac{3\pi}{8} + \frac{\pi^2}{8}\right) (l+l') X_0 D^3 +\nonumber\\ 
%&+& \frac{\pi}{2} l\,l' X_0^2 D^3 
v_{excl}(l,l', X_0) = 2 B_I X_0^2 \,l\, l' + 2 v_d k_I \frac{l+l'}{2}
\label{eq:vexcliso}
\end{equation}
where the parameters $B_I$ and $k_I$ can be estimated via MC integrals of a system composed by 
only two monomers (see Appendix~\ref{sec:appendix_a}) and $X_0$ is the aspect ratio of the monomers.
We assume that the chain length distribution $\nu(l)$ is exponential~\cite{ourMacromol} with an average chain length $M$ 
\begin{equation}
\nu(l) = \rho M^{-(l+1)} (M-1)^{l-1} %=\frac{\rho}{M (M-1)} e^{- l  [ \ln M - \ln (M-1) ]}
\label{eq:chaindistro}
\end{equation}
where
\begin{equation}
M = \frac{\sum_{l=1}^{\infty} l \, \nu(l)}{\sum_{l=1}^{\infty} \nu(l)}.
\label{eq:defM}
\end{equation}
With this choice for $\nu(l)$ the free energy in Eq. (\ref{eq:freeene}) becomes:
\begin{eqnarray}
\frac{\beta F_{I}}{V} &=&  -\rho \beta \Delta F_b (1 - M^{-1}) +\nonumber\\
&+& \eta(\phi) \left [ B_I X_0^2 + \frac{v_d k_I }{M} \right ]\rho^2  + \nonumber\\
&+&  \frac{\rho}{M} \left [ \ln\left ( \frac{v_d\rho}{M} \right )- 1\right ] + \nonumber\\
&+& \rho \frac{M-1}{M} \ln(M-1) - \rho \ln M.
\label{eq:Fiso}
\end{eqnarray}
%where, for example, for hard cylinders $B = D^3\pi/4$, $k_I= (3/2+\pi/2)\approx 3.071 $ and 
%$A= D^3  \pi^2/16$. 
%Note that in the present case $k_I$ and $B_I$ do not depend on $X_0$.

Minimization of  the free energy in Eq. (\ref{eq:Fiso}) with respect to $M$ provides the following expression for the average chain length $M(\phi)$:
\begin{equation}
M = \frac{1}{2} \left ( 1 + \sqrt{1 + 4 \phi e^{k_I \phi \eta(\phi) + \beta \Delta F_b} }\right).
\label{eq:avgchain}
\end{equation}

\subsubsection{Nematic phase}
In the nematic phase the monomer orientational distribution function  $f (\theta)$ depends explicitly on the angle $\theta$ between the particle and the nematic axis, i.e. the direction of average orientation of the DNAD,
since the system is supposed to have azimuthal symmetry around such axis.  We assume  the form proposed by Onsager~\cite{Onsager49}, i.e.:
\begin{equation}
f_{\alpha}(\theta) = \frac{\alpha}{4 \pi \sinh\alpha} \cosh(\alpha \cos\theta)
\label{eq:fons}
\end{equation}
where $\alpha$  controls the width of the angular distribution. Its equilibrium value is obtained by minimizing the 
free energy with respect to $\alpha$. 
%In view of the analytical expression for the excluded volume $v_{excl}$ for cylinders and for super quadrics~\cite{ourMacromol}, 
As discussed in Appendix~\ref{sec:appendix_a}, we assume the following form for the excluded volume in the nematic phase:
%, we assume the following form for the $v_{excl}$ of two DNADs averaged over the solid angle using the one parameter ($\alpha$) 
%dependent orientational distribution function $f_O({\bf u})$ defined in Eq. (\ref{eq:fons}):
\begin{equation}
%v_{excl}(l,l',\alpha) &=& 2 \left [ A_N(\alpha) + v_d k_N(\alpha) \frac{l + l'}{2} +\right .\nonumber\\ 
%&&\left .\phantom{\frac{l + l'}{2}}+ B_N(\alpha) X_0^2 l\,l' \;  \right ]   \hspace{0.5cm}
v_{excl}(l, l',X_0,\alpha)= 2 B_N(\alpha) X_0^2 l\, l' + 2 v_d k_N^{HC}(\alpha) \frac{l+l'}{2}
\label{eq:vexclnem}
\end{equation}
where the term $2 v_d k_N^{HC}(\alpha)$ is the end-midsection contribution to the excluded volume of two hard cylinders (see %Appendix~
Appendix~\ref{sec:appendix_b})  and 
\begin{equation}
B_N(\alpha) = \frac{\pi}{4} D^3 \left ( \eta_1 + \frac{\eta_2}{\alpha^{1/2}} + \frac{\eta_3}{\alpha} \right ).
\label{eq:BN}
\end{equation}
In Eq. (\ref{eq:BN}), $D$ is the diameter of the monomer and 
$\eta_k$ with $k=1,2,3$ are three parameters that we chose in order to reproduce the excluded volume calculated from MC calculations as discussed in %Appendix~
Appendix~\ref{sec:appendix_a}.

Inserting Eqs. (\ref{eq:vexclnem}) and (\ref{eq:chaindistro})
into Eq. (\ref{eq:freeene}) and assuming once more an exponential distribution for $\nu(l)$ one obtains  after some algebra:

\begin{eqnarray}
\frac{\beta F_{N}}{V} &=&\hat \sigma_o - \rho \beta \Delta F_b (1 - M^{-1}) +\nonumber\\ 
&+& \eta(\phi) \left [ B_N(\alpha) X_0^2 + \frac{ v_d k_N^{HC}(\alpha)}{M} \right ] \rho^2 +\nonumber\\
&+&  \frac{\rho}{M} \left ( \ln\left[\frac{v_d\rho}{M}\right ]-1\right ) 
- \rho\ln M + \nonumber\\
&+& \rho \ln(M-1) \frac{M-1}{M}\label{eq:Fnem}
\end{eqnarray} 
where $\hat \sigma_o \equiv \sum_l \sigma_o(l) \nu(l)$. 
The explicit calculation of the parameters $B_N$ and $k_N^{HC}$ is explained in Appendices
~\ref{sec:appendix_a} and~\ref{sec:appendix_b}.
% ~\ref{sec:appendix_a}
%and~\ref{sec:appendix_b}.

Assuming that the orientational entropy $\hat\sigma_o$
can be approximated with the expression valid for long chains~\cite{Odijk86}, minimization with respect to $M$  results in 
\begin{equation}
M = \frac{1}{2} \left ( 1 + \sqrt{1 + \alpha \phi e^{k_N(\alpha) \phi \eta(\phi) + \beta \Delta F_b} }\right).
\label{eq:avgchainnemL}
\end{equation}
while using the approximated expression for short chains~\cite{Odijk86}, one obtains
\begin{equation}
M = \frac{1}{2} \left ( 1 + \sqrt{1 + 4 \alpha \phi e^{k_N(\alpha) \phi \eta(\phi) + \beta \Delta F_b - 1} }\right).
\label{eq:avgchainnemS}
\end{equation}

The equilibrium value of $\alpha$ is thus determined by further minimizing the nematic free energy in Eq. (\ref{eq:Fnem}), which has become only a function of $\alpha$. 
 The parameter $\alpha$ is related to the degree of orientational ordering in the nematic phase as expressed by the nematic order parameter $S$
 %the nematic order parameter $S$ 
as follows:
 
\begin{equation}
S(\alpha) = \int  (3 \,\cos^2\theta-1) f_{\alpha}(\theta ) \pi \,\sin\theta \;d\theta \approx  1 - 3/\alpha. 
\end{equation}  
    
%In Ref.~\cite{ourMacromol} it has also been proposed an improved version of the free energy in the nematic phase
%(see Eqs. (40)-(42) therein), which accounts for the fact that the distribution of the monomer orientations at coexistence is nearly isotropic. 
Further refinements of the theory could be obtained by including a more accurate description of the orientational distribution $f_\alpha(\theta)$
in the proximity of the I-N phase transition, along the lines of Eqs. (40)-(42) of Ref.~\cite{ourMacromol}.
For the sake of simplicity we have just presented the basic theoretical treatment. However,
in the theoretical calculations in Sec.~\ref{sec:results} we will make use of the refined and more accurate free energy proposed in Ref.~\cite{ourMacromol}.

\subsubsection{Phase Coexistence}
The phase boundaries, at which the aggregates of DNAD are sufficiently long to induce macroscopic orientational ordering, are characterized by coexisting isotropic and nematic phases in which the volume fraction of DNADs are, respectively,
$\phi_N=v_d \rho_{N}$ and $\phi_I= v_d \rho_{I}$.
The number densities $\rho_I$ and  $\rho_N$ can be calculated by minimizing Eq. (\ref{eq:Fiso}) with respect to $M_I$ and by minimizing Eq. (\ref{eq:Fnem}) with respect to $M_N$ and $\alpha$. In addition, the two phases must be at equal pressure, i.e.   $P_{I}=P_{N}$, and chemical potential, i.e. $\mu_{I}=\mu_{N}$.
These conditions yield the following set of equations:

\begin{eqnarray}
\frac{\partial}{\partial M_{I}}  F_{I} (\rho_{I}, M_{I}) &=& 0\nonumber\\
\frac{\partial}{\partial M_{N}}  F_{N} (\rho_{N}, M_{N}, \alpha) &=& 0\nonumber\\
\frac{\partial}{\partial \alpha}  F_{N} (\rho_{N}, M_{N},\alpha) &=& 0\nonumber\\
P_{I}(\rho_{I}, M_{I}) &=& P_{N}(\rho_{N},M_{N},\alpha)\nonumber\\
\mu_{I}(\rho_{I}, M_{I}) &=& \mu_{N}(\rho_{N},M_{N},\alpha)
\label{eq:phasecoexeqs}
\end{eqnarray}

\section{Results and Discussion}
\label{sec:results}

\subsection{Properties of the model}
To characterize structural and geometrical properties of the simulated DNADs monomers and aggregates we analyze 
conformations of duplexes extracted from large-scale GPU simulations (see Figure~\ref{fig:snapshots} for some 
snapshots).

In the following, the volume $v_d$ occupied by a single DNAD of length $X_0 D$ and double helix diameter $D$
 ($D\simeq 2 \UOM{nm}$)
will be calculated as the volume of a cylinder with the same length and diameter, i.e. $v_d = \pi X_0 D^3 / 4$.
When comparing numerical and experimental results with theoretical predictions we use the number of nucleotides $N_b$ 
in place of $X_0$ ($X_0 \simeq 0.172 N_b$) and the concentration $c$ instead of the packing fraction $\phi$, which can be related to the former via:
\begin{equation}
\phi =  \frac{0.172 D^3\pi}{8 m_{N}}\, c
% \;\frac{8 m_{N}}{0.172 D^3 \pi}
\label{eq:cNphiN}
\end{equation}
\noindent
where $m_N=330 \UOM{Da}$ is the average mass of a nucleotide.
Hence, in the following $c_I$ and $c_N$ will be used in place of $\phi_I$ and $\phi_N$.

First we calculate the dimensions (height $L$ and width $D$) of the DNADs for different $c$ and $T$. 
We observe no concentration dependence on both quantities, while the variation in $T$ is negligible (of the order 
of $0.1\%$ between DNADs of samples at $270\UOM{K}$ and $300\UOM{K}$). The effect of this small change does not 
affect substantially the value of the aspect ratio, which we consider constant ($X_0=2.06$) throughout this work.

The geometrical properties of end-to-end bonded duplexes are not well-known since there are no experimental 
ways to probe such structures. In a very recent work, the interaction between duplex terminal 
base--pairs has been analyzed by means of large-scale full-atom simulations by Maffeo \textit{et al.}~\cite{DNAallatom12}. 
They found that blunt-ended duplexes (i.e. duplexes without dangling ends) have preferential 
binding conformations with different values of the azimuthal angle $\gamma$, defined as the angle between the 
projections onto the plane orthogonal to the axis of the double helix of the vectors connecting the O5' and 
O3' terminal base pairs. 
They report two preferential values for $\gamma$, namely $\gamma = -20^\circ$ and $\gamma = 180^\circ$. 

\begin{figure}[tbh]
\vskip 1cm
\includegraphics[width=.495\textwidth]{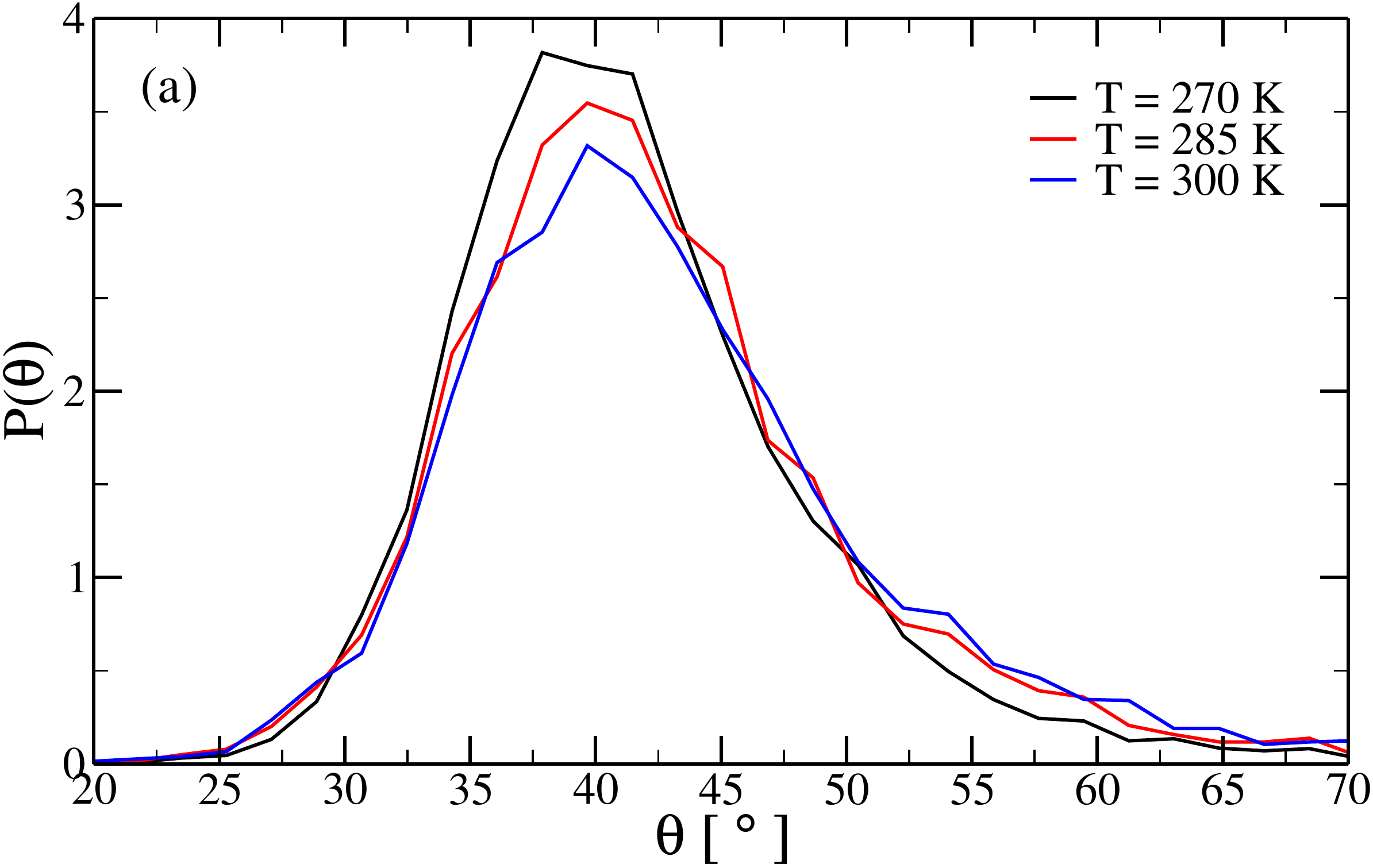}
\includegraphics[width=.495\textwidth]{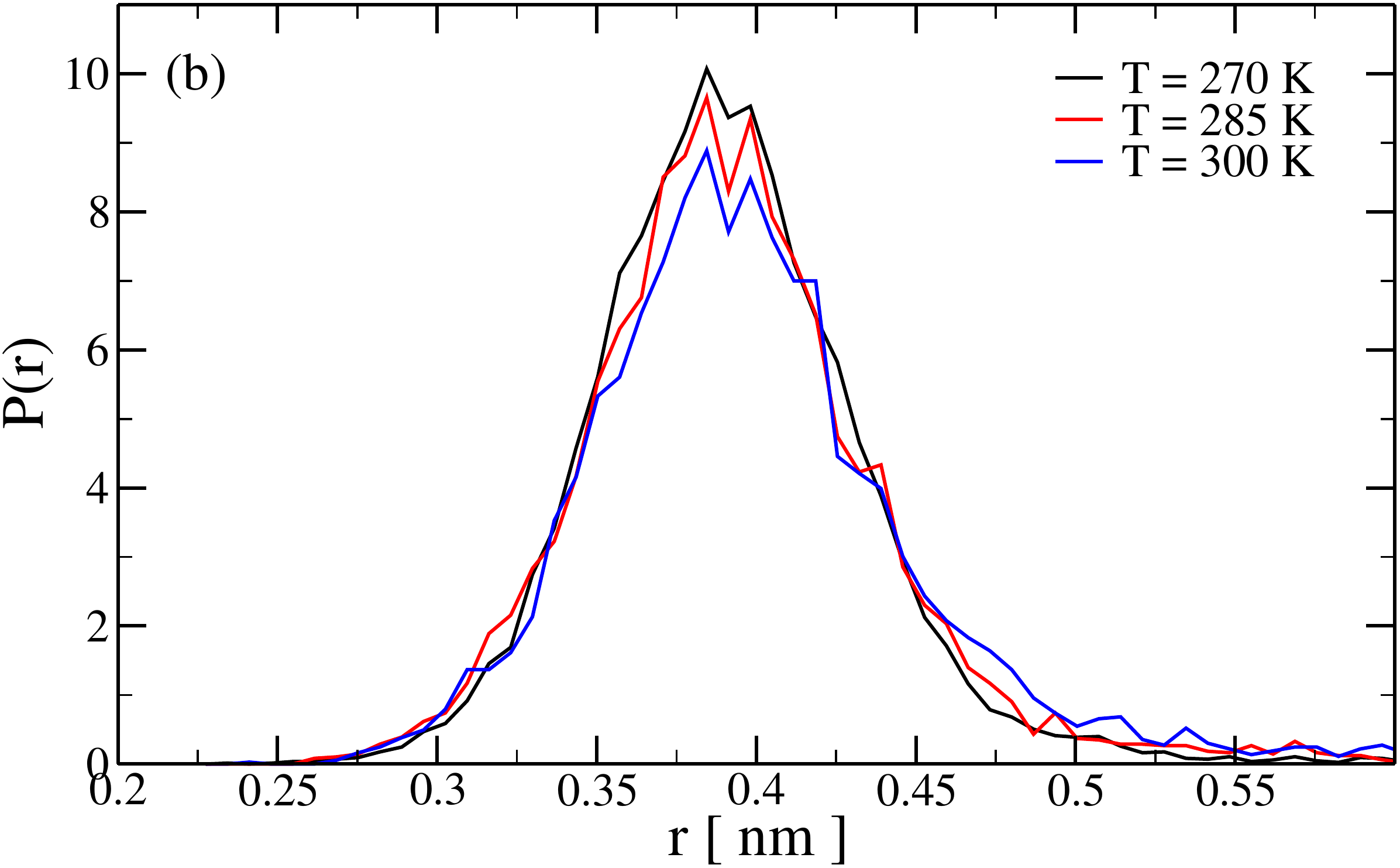}
\caption{Probability distributions for (a) the azimuthal angle $\gamma$ and (b) the end-to-end distance $r$.}
\label{fig:end_end}
\end{figure}

In the present model the continuity of the helix under end-to-end interactions is intrinsic in the model and the azimuthal 
angle probability distribution is peaked around a single value $\gamma_0 \approx 40^\circ$ (see Figure~\ref{fig:end_end}(a)). 
This is very close to the theoretical value $\gamma \approx 36^\circ$ given by the pitch of the B-DNA double helix. The 
qualitative difference between the conformations of bonded DNADs found in this work and in Ref.~ \cite{DNAallatom12}
should be addressed in future studies describing the coaxial end-to-end interaction in a more proper way.

In addition, we calculate the average distance $r$ between the centres of masses of the terminal base pairs. 
Figure~\ref{fig:end_end}(b) shows $P(r)$, the probability distribution of $r$. $P(r)$ is peaked at $0.39 \UOM{nm}$, 
whereas Maffeo \textit{et al.}~\cite{DNAallatom12} found an average distance of $r\approx 0.5 \UOM{nm}$. This difference can be 
understood in terms of the effect of the salt concentration which, being five times higher 
than the one used in  Ref.~\cite{DNAallatom12}, increases the 
electrostatic screening, thus effectively lowering the repulsion between DNA strands.

The effect of the temperature is small, as lowering $T$ leads only to more peaked 
distributions for both $P(\gamma)$ and $P(r)$ (and a very small shift towards smaller angles for $\gamma$) but 
does not change the overall behavior.

\subsection{Stacking free energy and excluded volume}
\label{sec:twobody}
%{\bf TOMMASO: qui l'intero paragrafo  un po' difficile da seguire. Mi chiedo se non possa tutto diventare una ulteriore appendice. Se rimane nel testo ci sarebbe bisogno di inserire forse qualche commento sul significato fisico ad esempio del risultato in figura 4.}

In this section we discuss the procedure employed to evaluate the input quantities required by the theory,
namely    $\Delta F_b$ and $v_{excl}(l,l')$. To this aim we perform a Monte Carlo 
integration over the degrees of freedom of two duplexes.
$\Delta F_b$ is defined as~\cite{ourMacromol}
\begin{equation} 
\beta \Delta F_b = \ln \left[2 \frac{\Delta(T)}{v_d}\right]  \label{eq:delfb} 
\end{equation}
where~\cite{BianchiJCP07}
\begin{equation}
\Delta(T) = \frac{1}{4}\left\langle\int_{V_b}  [ e^{-\beta V({\bf r}_{12},{\boldsymbol\Omega}_1,{\boldsymbol\Omega}_2)} - 1 ] \,d{\bf r}_{12} \right\rangle.%_{{\bf\Omega}_1,{\bf\Omega}_2} 
%\Delta(T) = 4 \pi \int_0^\infty g(r) \left\langle e^{-\frac{V(r)}{k_b T}} - 1 \right\rangle r^2 dr,
\end{equation}

Here ${\bf r}_{12}$ is the vector joining the center of masses of particles $1$ and $2$, ${\boldsymbol\Omega}_i$ is the orientation 
of particle $i$ and $\langle \ldots \rangle$ 
represents an average taken over all the possible orientations. $V_b$ is the bonding volume,
defined here as the set of points where the interaction energy $V({\bf r}_{12},{\boldsymbol\Omega}_1,{\boldsymbol\Omega}_2)$ between duplex $1$ and duplex $2$ is less
than $k_BT$. 
To numerically evaluate $\Delta(T)$ we perform a MC integration using the following scheme:

\begin{enumerate}
\item Produce an ensemble of 500 equilibrium configurations of a single duplex at temperature $T$.
\item Set the counter $N_{\mathrm{tries}} = 0$ and the energy factor $F = 0$.
\item Choose randomly two configurations $i$ and $j$ from the generated ensemble.
\item Insert a randomly oriented duplex $i$ in a random position in a cubic box of volume $V=1000$ $nm^{3}$. 
Insert a second duplex $j$ in a random position  and with a random orientation.
%which allows $i$ and $j$ to interact via coaxial stacking. 
%We defines as $V_{\rm avail}$ as the available volume in which $j$ can be placed.
%Choose a random orientation for $j$.
Compute the interaction energy $V(i,j)$ between the two duplexes $i$ and $j$ and, if $V(i,j)< k_BT$, update the energy factor, $F = F + \left(e^{-\beta V(i, j)} - 1\right)$.  Increment $N_{\mathrm{tries}}$.%{\bf USIAMO $V(\ldots)$ PER COERENZA CON LA DEFINIZIONE DI $\Delta$?}

\item  Repeat from step 3, until $\Delta(T) \cong \frac{1}{4} \frac{V}{N_{\rm tries}} F$ converges within a few per cent precision.
\end{enumerate}

The employed procedure to compute $v_{excl}(l,l')$ is fairly similar except that it is performed for
duplexes with a various number of bases (i.e. with different $X_0$) and the quantity $F$ counts how many trials originate a pair configuration with $V(i,j)> k_BT$ (i.e. in step $4$, $F = F+1$). In the nematic case, the orientations of the duplexes are extracted randomly from the Onsager 
distribution given by Eq.~\ref{eq:fons}. With such procedure,
 
\begin{equation}
v_{excl}(l=1, l'=1,X_0) = \frac{V}{N_{\rm tries}} F
\end{equation}

We calculate $v_{excl}$ for $8$ values of $\alpha$, ranging from $5$ 
to $45$ (see Appendix~\ref{sec:appendix_b}).  Since the $X_0$ and $l$ dependences of Eqs.~(\ref{eq:vexcliso}) and~(\ref{eq:vexclnem}) are the same and the $X_0$ dependence of the numerically calculated $v_{excl}$ on the shape of DNADs is negligible, the evaluation of the excluded volume as a function of
$X_0$ provides the same information as  the evaluation of $v_{excl}$ as a function of $l$.

\begin{figure}[tbh]
\vskip 1cm
\includegraphics[width=.495\textwidth]{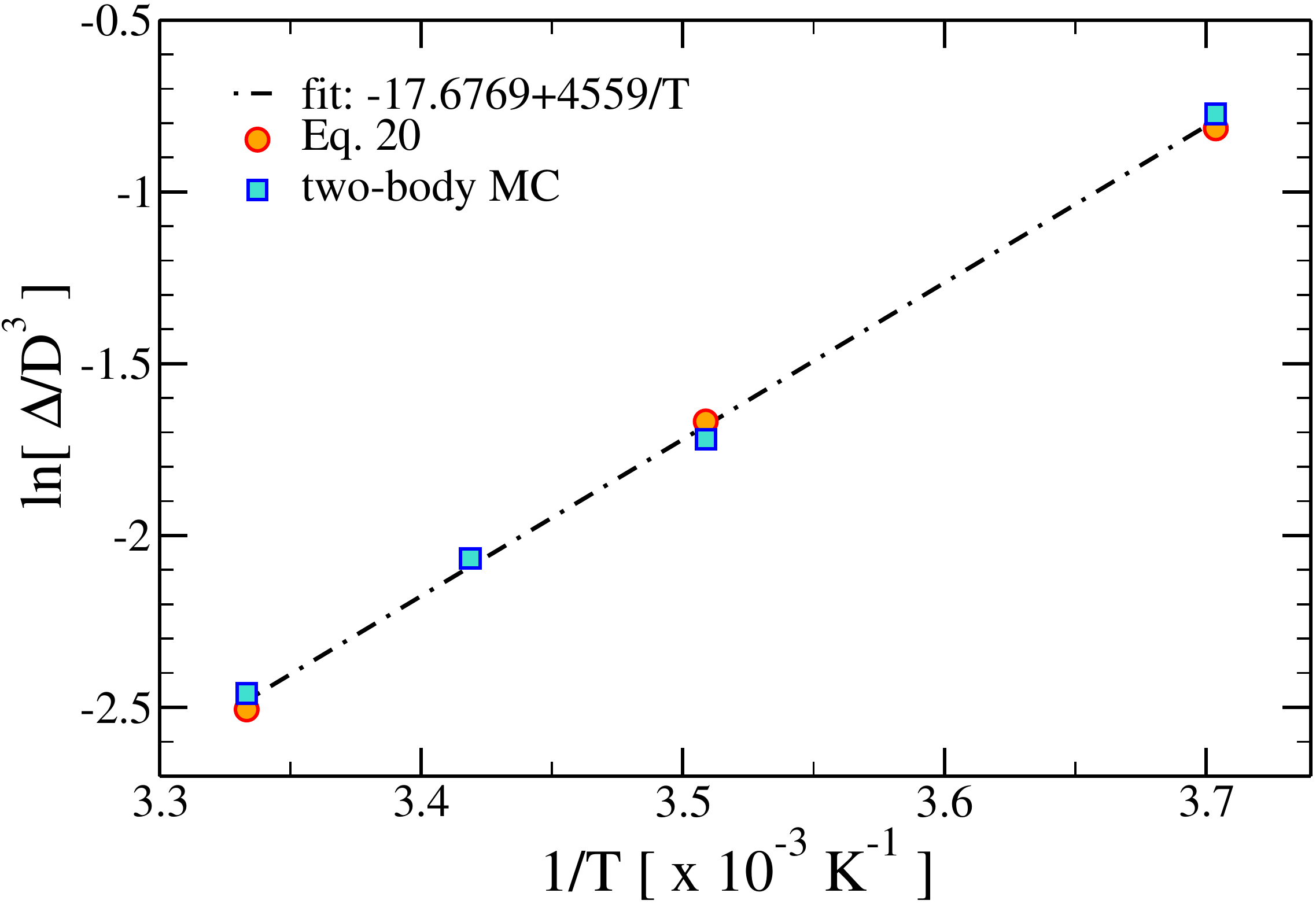}
\caption{$\Delta(T)$ calculated with the procedures described in Sec.~\ref{sec:twobody} for all investigated $T$.}
\label{fig:calcdelT}
\end{figure}

Fig.~\ref{fig:calcdelT}  shows  $\Delta(T)$ for all investigated $T$   in a  $\ln \Delta $ vs $1/T$ plot.
A linear dependence properly describes the data at the three $T$. 
An alternative way to evaluate $\Delta(T)$ is provided by the limit $\rho \to 0$ of Eq.~(\ref{eq:avgchain}).
Indeed in the low density limit $M$ and $\Delta(T)$ are related via the following relation:
\begin{equation}
\Delta(T) =  \frac{M (1- M)}{2\rho}.
\end{equation}

Therefore it is also possible to estimate $\Delta(T)$ by extrapolating the low density data for $M$ at $T=270\UOM{K}$, 
$285\UOM{K}$ and $300\UOM{K}$. The results, also shown in Fig.~\ref{fig:calcdelT}, are in line with the ones obtained 
through MC calculations.
The Arrhenius behavior of $\Delta(T)$ suggests that bonding entropy and stacking energy are in first approximation $T$ independent.
%Substituting the fit expression in the following equation relating the coaxial stacking free energy $G_{ST}$ to
%$\Delta(T)$
The coaxial stacking free energy $G_{ST}$ is related to $\Delta(T)$ as follows

\begin{equation}
G_{ST}= -  k_B T \ln [ 2 \rho \Delta(T) ].
\label{eq:del2Gst}
\end{equation}
\noindent Substituting the fit expression provided in Fig. \ref{fig:calcdelT} for $\Delta(T)$
results in a stacking free energy $G_{ST}^0=-0.086 %-0.17
\UOM{kcal/mol}$  at  a standard concentration
$1 \UOM{M}$ of DNADs and $T=293\UOM{K}$ comprising  a bonding entropy of $-30.6$ %$-61.2$ 
$\hbox{cal}/\hbox{mol}\,\hbox{K}$ and a bonding energy of $-9.06 %-18.1
\UOM{kcal/mol}$.
%Unfortunately a direct experimental quantification of $G_{ST}$ can not be found in literature. Several experimental techniques~\cite{NuclAcResGST06,NuclAcResGST01,NuclAcResGST00,JMolBiolGST04,ivanovaGST03}  have been used to measure quantities somehow related to the coaxial stacking free energy of binding, but a direct comparison with the present estimate of $G_{ST}$ can not be carried out. 

%In the study carried out by Maffeo {\it et al.} $G_{ST}$ has been estimated to be $-6.3 \UOM{kcal/mol}$, a value
%somewhat smaller than ours 

%We note that the experiments considered to parametrize the coaxial stacking interactions
%quantify the stacking interactions by observing the difference in the mobility of a double strand, where one of the 
%two strands has been nicked, relative to intact DNA~\cite{JMolBiolGST04,NuclAcResGST06}. These experiments are qualitatively 
%different from the ones taken into account in the present work (i.e. self-assembly of double-stranded DNA duplexes) especially 
%in the way they account for the entropic contribution to the stacking free energy.

\subsection{Isotropic phase: comparing simulation results with theoretical predictions }
\begin{figure}[tbh]
\vskip 1cm
\includegraphics[width=.495\textwidth]{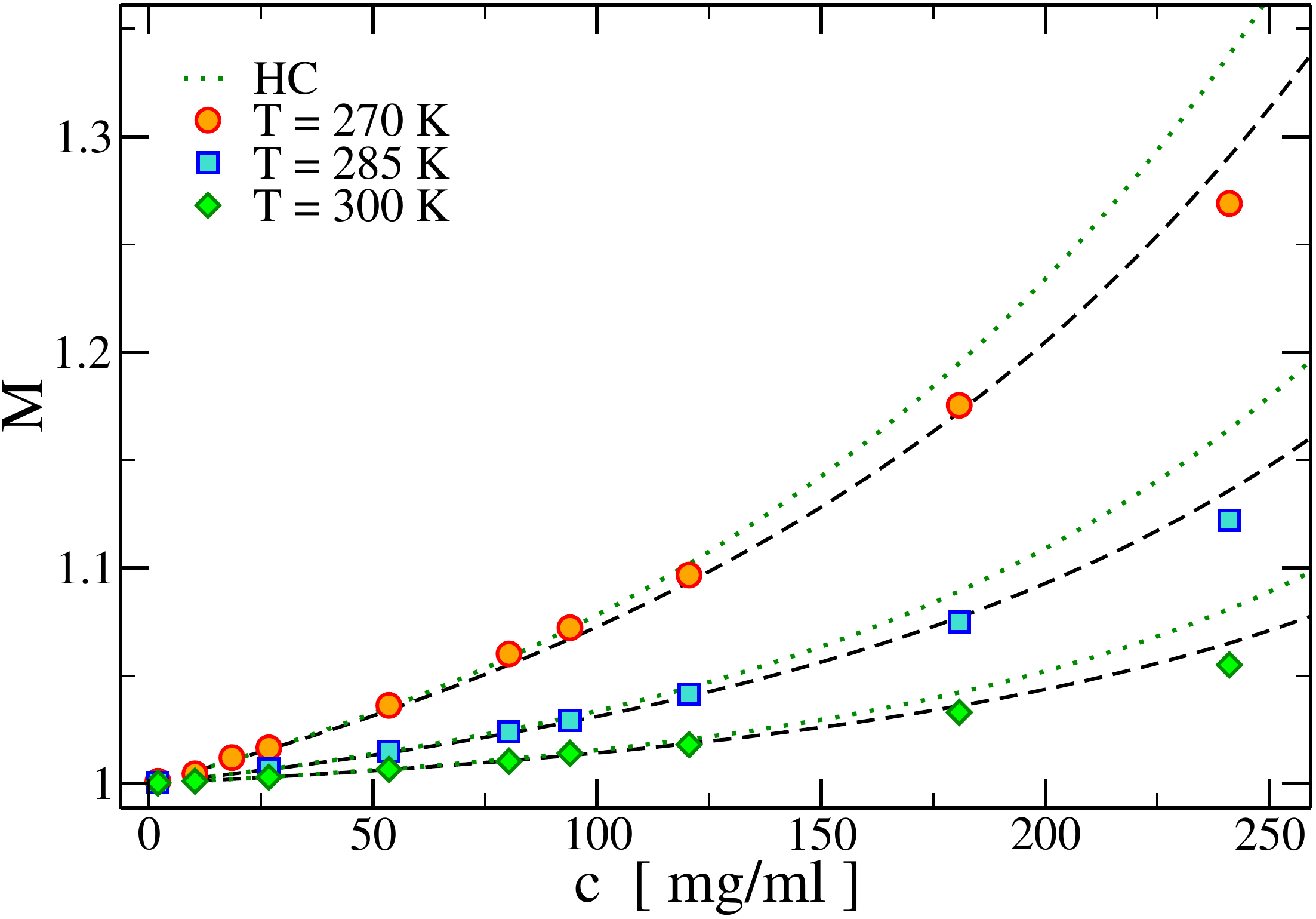}
\caption{Average chain length $M$ in the isotropic phase at low concentration. Symbols are numerical
results and dashed lines are theoretical predictions. Dotted lines are theoretical predictions using 
%in Eqs. (\ref{eq:Fiso}) and (\ref{eq:Fnem}) 
the excluded volume of HCs  $v_{excl}^{HC}$ (see Appendix~\ref{sec:appendix_b}).}
\label{fig:lmedio}
\end{figure}

Fig.~\ref{fig:lmedio} shows the concentration $c$ dependence of  $M$   calculated from the MD simulation of  the 
$N_b=12$  system.  The average  chain length increases progressively on increasing $c$.
The figure also shows the theoretical predictions calculated by minimizing the isotropic free energy in Eq.  (\ref{eq:Fiso}) with respect to $M$ using  the previously discussed estimates for  $\Delta F_b$ and $v_{excl}$. 
The theoretical results properly describe the MD simulation data  up to concentrations around $200\UOM{mg/ml}$,
which corresponds to a volume fraction $\phi \approx 0.20$. In Ref.~\cite{ourMacromol} similar observations 
have been made and the discrepancy  at moderate and high $\phi$ has been attributed  to the inaccuracy of the Parsons decoupling approximation. The $M$ values calculated using the excluded volume of
two hard cylinders (HC) are also reported, to quantify the relevance of the actual shape of the DNA duplex. 
Indeed the HC predictions appreciably deviate from numerical data beyond $100$ mg/ml.
 
%Up to volume fractions around $\phi\approx0.20$ the agreement between theoretical and numerical results
%is quite good for all cases considered. Above this volume fraction the theoretical predictions start
%deviating appreciably, a discrepancy that we attribute at
%moderate and high $\phi$ to the inaccuracy of the Parsons decoupling approximation.   
%We also plot in Figs.~\ref{fig:MforAll} (a)-(c) as dotted lines the predictions based on a Onsager-like 
%theory, i.e. setting the Parsons-Lee factor $\eta(\phi)$ equal to $1$ in Eq.(\ref{eq:freeene}). At low
%volume fractions the  approximation   $\eta(\phi)=1$ does not affect the quality of the results but above
%$\phi\approx0.20$ the use of Parsons decoupling approximation seem to better capture the behavior of $M(\phi)$. 
% <<< OLD 22/08
%In Fig.~\ref{fig:MforAll} (d) we report both the numerical and the theoretical  --- calculated according to Eq. (\ref{eq:Fiso}) with $M$ obtained by minimization of the isotropic free energy ---
%aggregate size distribution $\nu(l)$. 
%===
%In Fig.~\ref{fig:MforAll} (d) we report the aggregate size distribution $\nu(l)$ as obtained from both simulation and  
%theory, the latter calculated according to Eq. (\ref{eq:Fiso}) with $M$ obtained by minimization of the isotropic free energy. 
% >>>
%As expected, the aggregate size  
%in  the isotropic phase is  exponential.     These results suggest that a reasonable first principles
%description of the isotropic phase is provided by the free energy of Eq.~(\ref{eq:Fiso}), when the
%parameters of the model are properly evaluated.  

\subsection{Phase Coexistence: Theoretical predictions}

\begin{figure}[tbh]
\includegraphics[width=.495\textwidth]{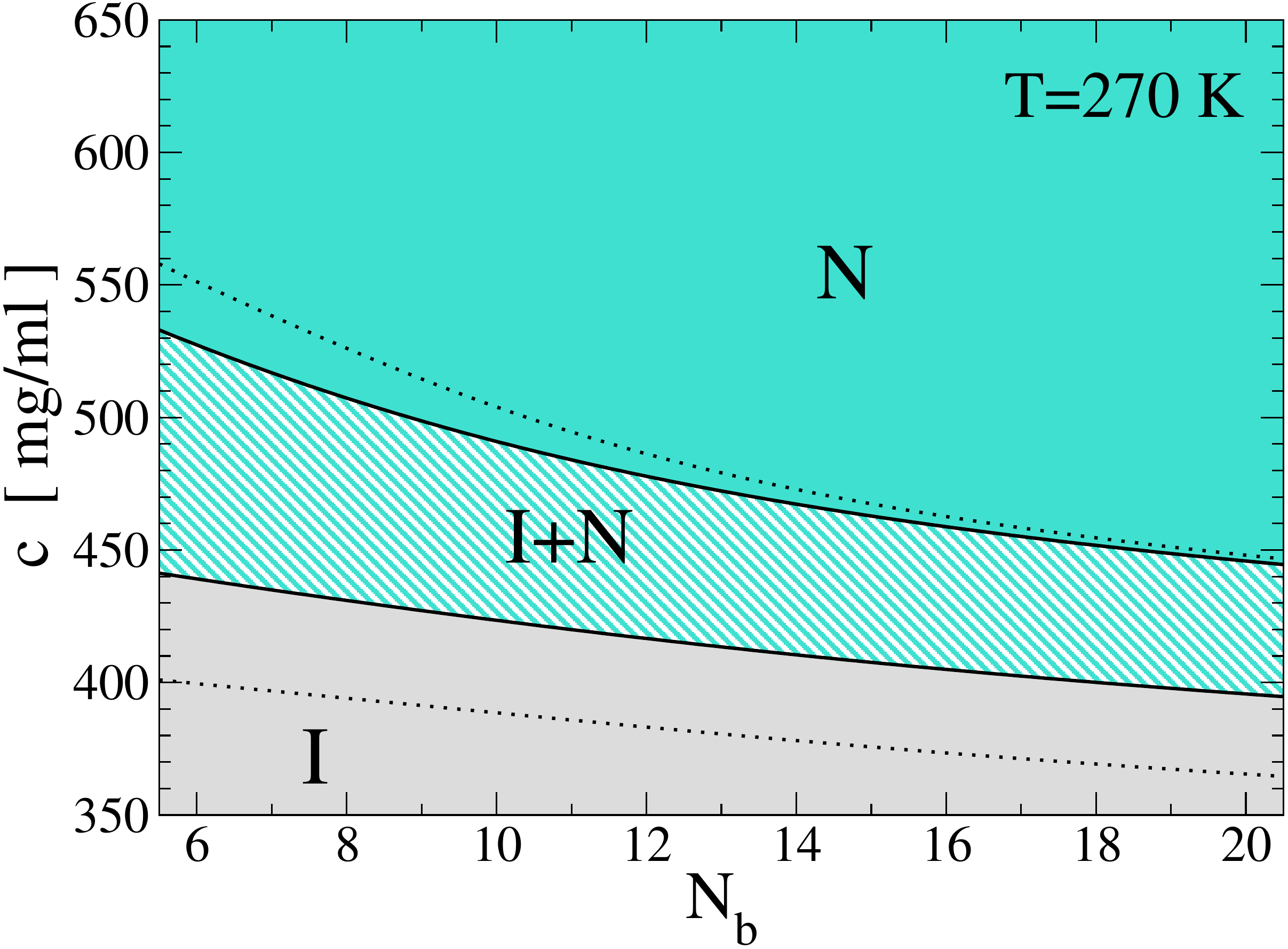}
\includegraphics[width=.495\textwidth]{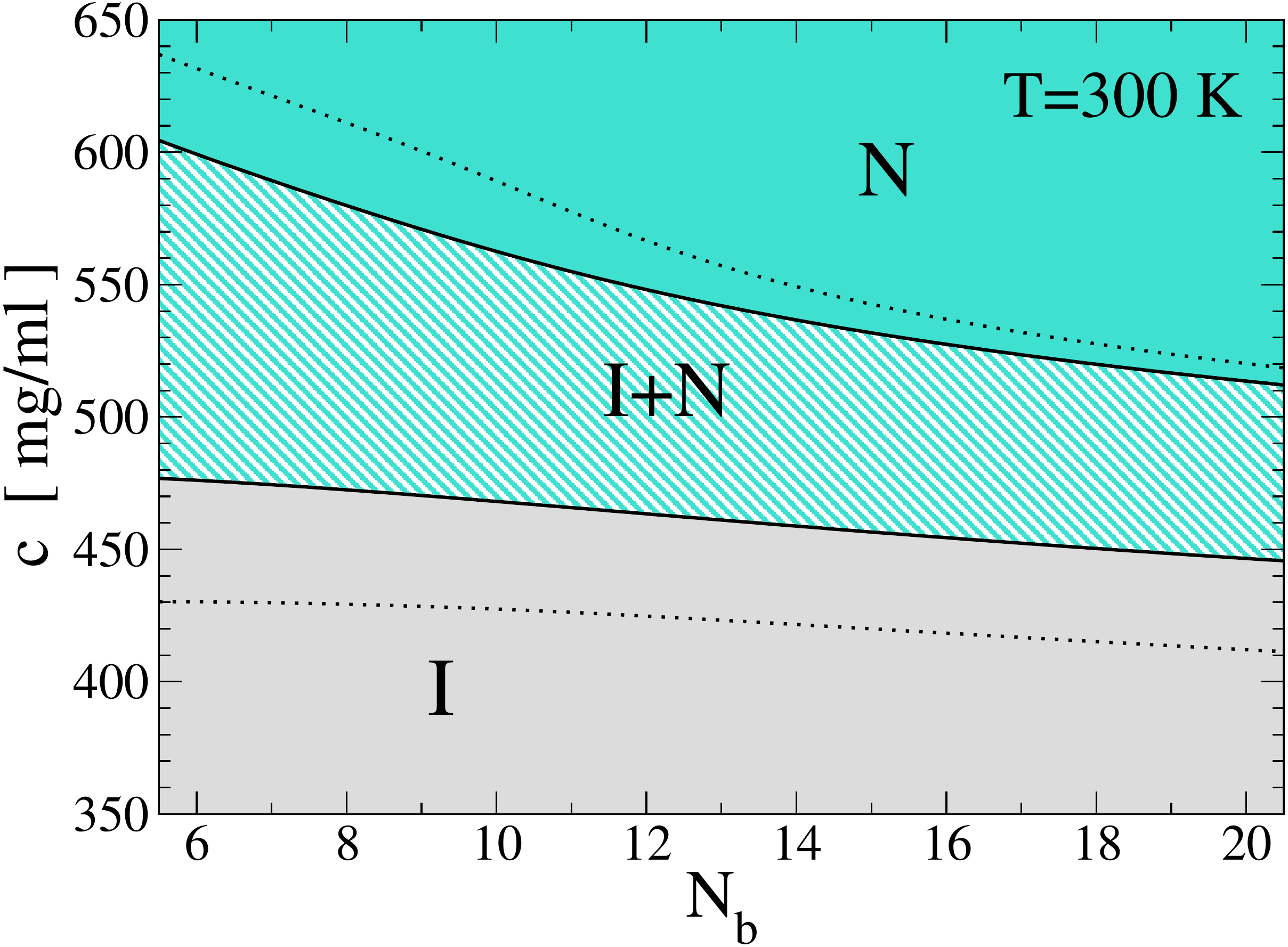}
\caption{I-N phase diagram in the $c$ vs $N_b$ plane for $T=270 \UOM{K}$ (top) and $300 \UOM{K}$ (bottom). Dotted lines
are theoretical phase boundaries calculated using the excluded volume of HCs $v_{excl}^{HC}$ (see Appendix~\ref{sec:appendix_b}).}
\label{fig:phasediag}
\end{figure}

A numerical evaluation of the phase coexistence between the isotropic and the nematic phases   for the coarse-grained model adopted in this study is still impossible to obtain given the current computational power.  
%out of present day numerical resources. 
We thus limit ourselves to the evaluation of the I-N phase coexistence via the theoretical approach 
discussed in Sec.~\ref{sec:theory}. Fig.~\ref{fig:phasediag} shows  the theoretical phase diagram in the $c$-$N_b$ plane
for $T=270\UOM{K}$ and  $300\UOM{K}$.   As expected, both  $c_I$ and  $c_N$ decrease on increasing $N_b$, since
the increase of the number of basis result in a larger aspect ratio.  On decreasing $T$, theory predicts  a $10\%$ decrease of  $c_I$  and a similar decrease of $c_N$, resulting in an overall shift of the I-N coexistence region toward lower $c$ values.  This trend 
%can be interpreted as resulting from 
is related to the increase of the average chain length  $M$  with increasing $\beta \Delta F_b$ (see Fig.~\ref{fig:calcdelT}).

%The theoretical values for the average chain length at the nematic-isotropic coexistence are
%shown in Fig.~\ref{fig:MforAll}(a)-(c). Along the $\phi_I$ transition line, $M$
%ranges from $2$ to $4$. On the contrary along the $\phi_N$ transition line, the $M$ values are larger   
%and depend on aspect ratio and stacking energy.

Fig.~\ref{fig:phasediag} also shows the phase boundaries calculated using the excluded volume of two hard cylinders.  Assimilating DNADs to hard cylinders  results in a $10$--$15 \%$   widening of the  isotropic-nematic coexistence region.

\subsection{Comparison between theory and  experiments}
%%we selected the lowest transition concentration among the ones experimentally determined, since these would be closest to the symmetric %%monomers considered in the model. 

The theoretical predictions  concerning the isotropic-nematic  coexisting concentrations can be  compared to the experimental results reported in Refs.~\cite{BelliniScience07} and~\cite{BelliniPNAS2010} for blunt-ended DNADs.
 
\begin{figure}[tbh]
\vskip 1cm
\includegraphics[width=.495\textwidth]{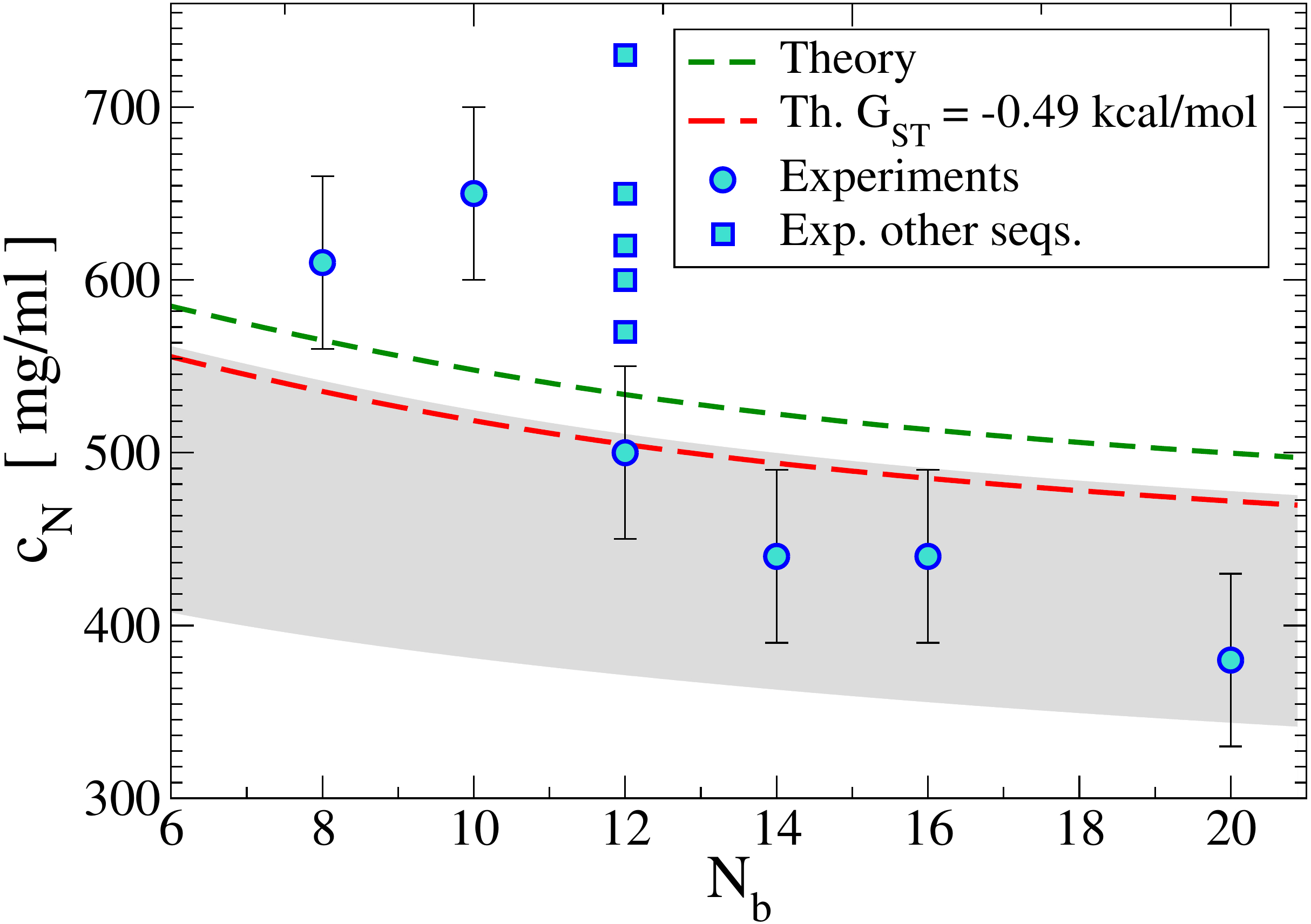}
\caption{Critical nematic concentrations $c_N$ as a function of the number of base pairs per duplex $N_b$ for the 
present model, calculated theoretically at $T= 293 \UOM{K}$ using the computed stacking free energy $G_{ST}^0$ %shown in Fig.~\ref{fig:calcdelT} 
  (short dashed lines), $G_{ST}=-0.49 \UOM{kcal/mol}$  (long dashed lines), and for experiments~\cite{BelliniScience07} (circles and squares). Squares are $c_N$ for different sequences at the same $N_b = 12$.
The grey band has been built considering  for $G_{ST}$ %experimental
an upper bound of $-0.4 \UOM{kcal/mol}$ and a lower bound of $-2.4 \UOM{kcal/mol}$. }
\label{fig:compexp}
\end{figure}

%These experimental data can be transformed into volume fractions once the relevant properties of DNAD are known
%(DNAD molecular weight $m_{D}=660\,N_b\; \hbox{Da}$,  diameter $D \approx$ $2\; nm$,  length $L = N_b/3\; nm$, where $N_b$ is the number of bases in the sequence). 
%The number density $\rho$ of DNADs is related to the  mass concentration
%\begin{equation} 
%\rho = \frac{c}{m_{D}} 
%\end{equation}  
%Since $v_{d}=L D^2 \pi / 4 $ is the volume of a DNAD,  the volume fraction can be expressed as:
%\begin{equation} 
%\phi = \rho v_{d} = \frac{c L D^2 \pi}{4 m_{D}} 
%\label{eq:c2phi} 
%\end{equation}  
Figure~\ref{fig:compexp} compares the experimentally determined  nematic concentrations $c_N$ at coexistence
with the values calculated from the present model for  $T=293 \UOM{K}$. %and $T=300 \UOM{K}$.  
Despite all the simplifying assumptions and despite the experimental uncertainty, the results provide a reasonable 
description of the $N_b$ dependence of $c_N$.
%We also report the values calculated from the SQ model for  $\beta \Delta E_S=6.67$ and $\beta\Delta E_S=5.56$.
%As in Ref.~\cite{ourMacromol}, we show the lowest experimental values for the transition concentration,  since in these systems the DNADs geometry should be the most similar (i.e. straight as opposed to bended) to the present model.
The experimental data refer to different base sequences and different salt concentrations. According to the authors $c_N$ is affected by an error of about $\pm 50 \UOM{mg/ml}$.  In particular  for the case $N_b=12$ the critical concentrations $c_N$ for distinct sequences shows  that blunt-end duplexes of equal length but different sequences can display significantly different transition concentrations.
Hence, for each duplex length, we consider the lowest transition concentration among the ones experimentally determined, since this
corresponds to the sequence closest to the symmetric monomer in the model. 
Indeed the dependence of  $c_N$ on the DNADs sequence is expected to be larger for the shortest sequences, i.e. $N_b < 12$,  for which DNAD bending could be significant~\cite{BelliniPNAS2010}. Unfortunately, quantitative experimental data on this bending effect are still lacking. %We would expect this to be particularly true for the shortest sequences, in which the effect of bent helices could be more relevant.
In general it is possible that $c_N$ for $N_b < 12$ (for which a large number of sequences have been studied, see Fig.~\ref{fig:compexp}),  would be corrected to lower values if a larger number of sequences were explored. 
For more details on this phenomenon, we refer the reader to the discussions in Refs.~\cite{BelliniPNAS2010,FerrariniSM2011,ourMacromol}.

The overestimation  of the phase boundaries  for $N_b \geq 12$ with respect to experimental results suggests that the DNA model of Ouldridge {\it et al.}~\cite{ouldridge_jcp}  overestimates the coaxial stacking free energy.
%er than previous estimates.  
Such discrepancy  can perhaps be attributed to the restricted number of microstates allowing for bonding states in the DNA model~\cite{ouldridge_jcp,tesi_tom}, as discussed in Sec.~\ref{sec:results}. 
Indeed, allowing DNADs to form end-to-end bonds with more than one preferred 
azimuthal angle would increase the entropy of bonding, thus effectively lowering $G_{ST}$. 
Allowing for both left- and right-handed 
binding conformations, a possibility supported by the  results of Maffeo {\it et al.}~\cite{DNAallatom12}, would double $\Delta(T)$ in Eq.~(\ref{eq:del2Gst}) and hence add an entropic contribution equal to $-k_B T\ln(2)$ to $G_{ST}$,   
which would result in  a  value $G_{ST}^0 - 0.403\UOM{kcal/mol} = -0.49 \UOM{kcal/mol}$ for $T=293K$.
Fig.~\ref{fig:compexp} also shows the theoretical prediction for such upgraded $G_{ST}$ value.  

In Fig.~\ref{fig:compexp} theoretical calculations of the I-N transition lines are shown for $G_{ST}=-0.4 \UOM{kcal/mol}$ and $G_{ST} =-2.4 \UOM{kcal/mol}$ at $T=293 \UOM{K}$ as the upper and lower boundaries of the grey band respectively.
To calculate these critical lines we retain the excluded volume calculated in the subsection~\ref{sec:twobody} 
and, given the value of $G_{ST}$, we evaluate $\Delta F_b$ according to Eqs. (\ref{eq:delfb}) and (\ref{eq:del2Gst})
for $T=293 \UOM{K}$ and $\rho$ corresponding to the standard $1 \UOM{M}$ concentration.

The selected points with $N_b \geq 12$ fall within the grey band shown in Fig.~\ref{fig:compexp} 
 enabling us to provide an indirect estimate of $G_{ST}$ between $-0.4 \UOM{kcal/mol}$ and 
$-2.4 \UOM{kcal/mol}$.
For the points with $N < 12$, where duplex bending might play a role, it would be valuable to have more
experimental points corresponding to more straight sequences in order to validate the theoretical predictions.
   
%In Fig.~\ref{fig:compexp} is also shown a grey band marking the lower and upper bounds for $G_{ST}$

%Finally it is worth observing that for $N_b > 18$ the DNADs critical concentration can not be high enough to ensure a
%counter ions concentrations which are able to fully screen the electrostatic interactions between the charged DNADs~\cite{BelliniPNAS2010}. For that reason the point $N_b=20$ shown in Fig.\ref{fig:compexp} has been obtained adding to the water solution salt with a $1.2$ $M$ concentration.

It is worth observing that for all DNAD lengths $N_b$ the electrostatics interactions are properly screened. For $N_b=20$
a concentration $1.2 \UOM{M}$ of NaCl has been added to the solution resulting in a Debye screening length $k_D^{-1} \approx 0.23 \UOM{nm}$. For all other lengths (i.e. $N_b \leq 18$)  we note that at the lowest DNA concentration of $440 \UOM{mg/ml}$ 
corresponding to $N_b=14$ $k_D^{-1} \approx 0.40 \UOM{nm}$.
Therefore the experimental $k_D^{-1}$ is always smaller than the excluded volume diameter for the backbone-backbone interaction
of our coarse-grained model~\cite{ouldridge_jcp} ($ \approx 0.6 \UOM{nm}$), thus enabling us to neglect electrostatic interactions.

On the other hand, if electrostatics interactions are not properly screened  the effective aspect ratio for such DNAD sequences would be smaller than the ones used in our theoretical treatment and this would result in a underestimate of $c_N$.  
To account for this behavior one should at least have a reasonable estimate of the effective size of DNADs when electrostatics
interactions are not fully screened. Moreover, the role of electrostatics interactions can be subtle and not completely accounted for by simply introducing an effective size of DNADs. A possible route to include electrostatics in our treatment can be found in Ref.~\cite{FerrariniJCP05} and it will be addressed in future studies.

\subsection{Comparison with Onsager Theory}
\begin{figure}[tbh]
\vskip 1cm
\includegraphics[width=.495\textwidth]{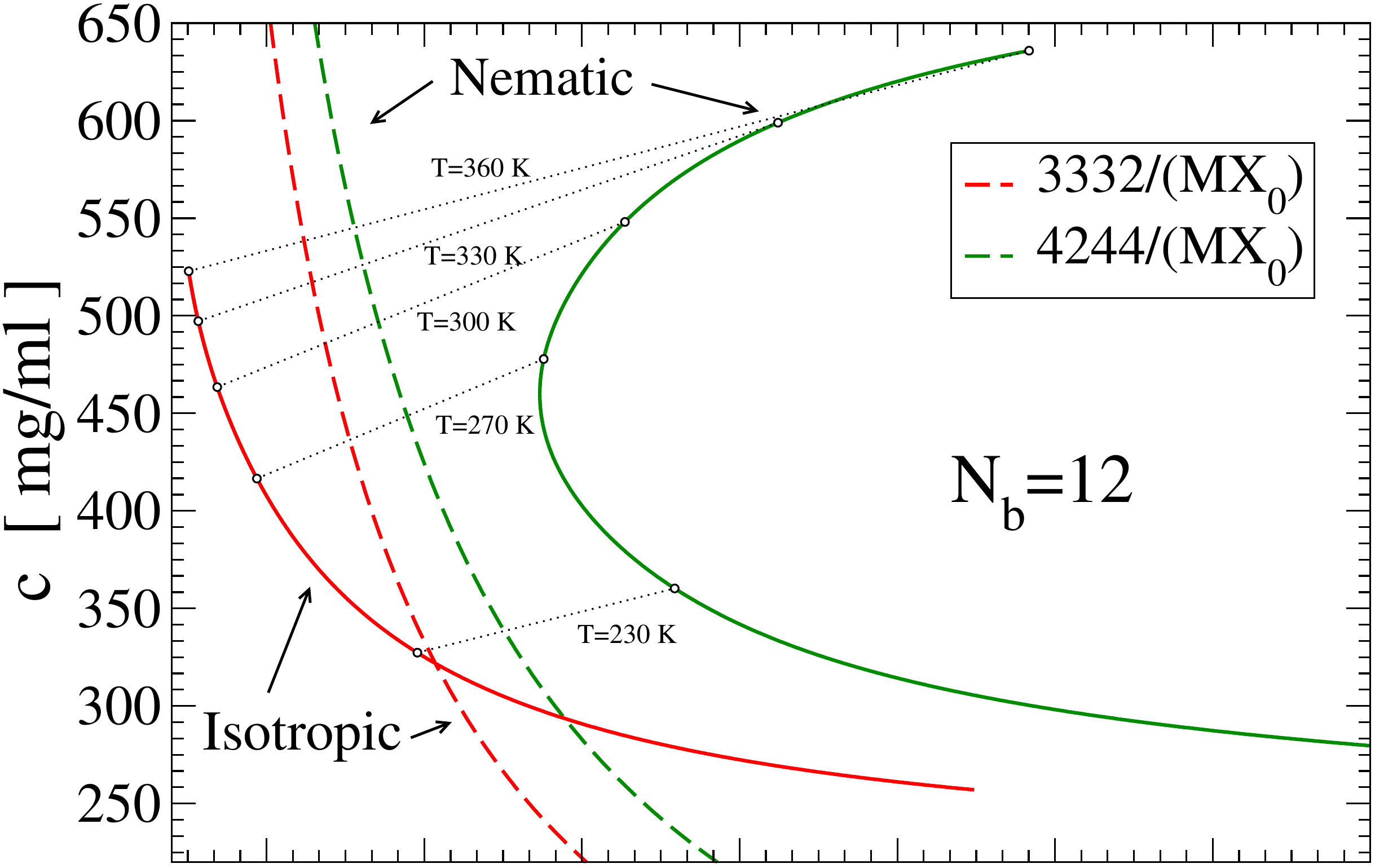}
\includegraphics[width=.504\textwidth]{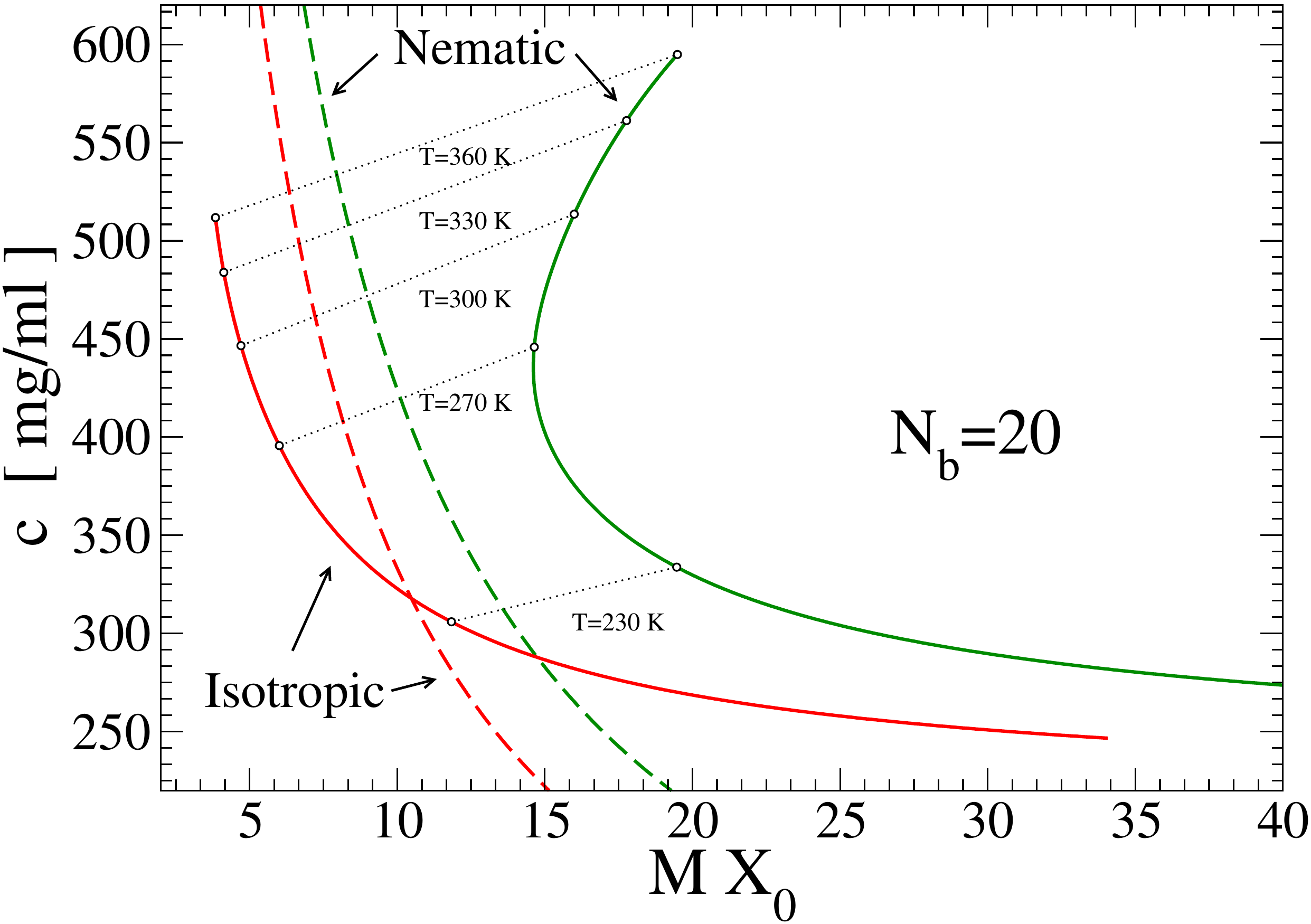}
\caption{
Isotropic-nematic coexistence lines in the average aspect ratio $M X_0$  and concentration $c$ plane for two values of $N_b$, namely $N_b=12$ (top) and $N_b = 20$ (bottom). Solid lines indicate theoretical predictions, dashed lines indicate the  Onsager original predictions, as re-evaluated 
in Ref. ~\cite{Vroege92} for $c_{I}$ and $c_N$.  Symbols along the isotropic and nematic phase boundaries at coexistence 
are joined by dotted lines, to indicate the change in concentration and average chain length at the transition.}
\label{fig:re-entr}
\end{figure}

The experimental average aggregation numbers are  estimated in Refs.~\cite{MezzengaLangmuir2010,BelliniScience07} by mapping the self-assembled system onto 
an ``equivalent'' mono-disperse system of hard rods with an aspect ratio equal to $M X_0$.
In Ref.~\cite{ourMacromol} it has been shown that the theoretically estimated isotropic-nematic coexistence lines for the case of polymerizing superquadric particles 
in the $M X_0-\phi$ plane, parametrized by the stacking energy, are significantly different from the corresponding Onsager  original predictions (as re-evaluated in Ref.~\cite{Vroege92}).  In light of the relevance for interpreting the experimental data,
we show in  Figure~\ref{fig:re-entr} the same curves for the DNA model investigated here.
In this model,  a clear re-entrant behavior of the transition lines in the $c-M X_0$ plane is observed. The  re-entrant behavior occurs  for values of the stacking free energy accessed   at temperatures  between $270 \UOM{K}$ and $330 \UOM{K}$. 
We believe that %, although a quantitative agreement between our theory and experiments is not to be expected,  
the re-entrancy of the transition lines in the $c-M X_0$ plane is a peculiar mark of the system polydispersity  resulting from the reversible self-assembling into chains, and as such it should be also observable experimentally.
 
\section{Conclusions}
\label{sec:conclusions}

In this article, we have provided the first  study of a bulk solution of blunt ended DNA
duplexes undergoing reversible self-assembly into chains, promoted by %by exploiting coaxial 
stacking interactions.  The simulation study, carried out at different concentrations
and temperatures, provides a clear characterization of the $c$ and $T$ dependence of
the average polymerization length $M$ and an indirect estimate of the
stacking free energy.   We have provided a theoretical description
of the self-assembly process based on a  theoretical framework recently developed in Ref.~\cite{ourMacromol}. The inputs required by the theory (the DNAD excluded volumes and the stacking free energy) have been numerically calculated for the present DNA model, allowing a parameter free comparison between the molecular dynamic results and the theoretical predictions. Such  comparison 
has been limited to the isotropic phase, due to the difficulties to simulate 
 the dense nematic phase under equilibrium conditions. The description of the isotropic phase is satisfactory: quantitative agreement  between theory and simulations is achieved for concentrations up to $c\approx 200$ mg/ml. The stacking free energy value that properly accounts for the polymerization process observed in the molecular dynamics simulations  is $G^0_{ST}=-0.086\UOM{kcal/mol}$  at a standard concentration 1 M of DNADs %{\bf 1 M OF WHAT ?? SINGLE HELIXES ? DOUPLEXES ? BASES ?} 
and $T = 293~$K comprising a bonding entropy of $-30.6 \UOM{cal/mol~K}$ and a bonding energy of $-9.06 \UOM{kcal/mol}$.	

Theoretical predictions for the I-N transition have been compared with experimental results
for several DNA lengths, ranging from 8 to 20 bases.   For $N_b \geq 12 $
% where experiments are de-facto performed in high electrostatic screening conditions,  
the model predicts values for $c_N$ which are higher than experimental ones. This suggests that the DNA model employed overestimates $G_{ST}$.  In view of the recent results of Maffeo {\it et al. }\cite{DNAallatom12}, we speculate that the bonding entropy  is underestimated, in agreement with the observation that  the probability distribution of the azimuthal angle between two bonded DNADs, which is designed to be single-peaked, is too restraining. In this respect, the present study call for an improvement of the coarse-grained potential~\cite{ouldridge_jcp} in regard to the coaxial stacking interaction.

The value of $G_{ST}$ can also be used as a fitting parameter in the theory 
for matching  $c_N$ with the experimental results, retaining the
excluded volume estimates calculated for the coarse-grained DNA model.  
Such procedure shows that  values of the stacking free energy between $-0.4\UOM{kcal/mol}$ and $-2.4\UOM{kcal/mol}$ are 
compatible with the experimental location of the I-N transition line.  
 In the work of Maffeo \textit{et al.}, the authors report a quite smaller value of $G_{ST}$, namely  $G^M_{ST}=-6.3\UOM{kcal/mol}$, a value which was confirmed  by the same authors by  performing an investigation of the aggregation kinetic in a very lengthy all-atom simulation  of DNAD with $N_b=10$.  If such $G_{ST}$ value is selected as input in our theoretical approach (maintaining the same excluded volume term),  then one finds $c_N^M \approx 250 $ mg/ml, a value significantly smaller than the experimental result ($c_N=650\pm 50$ mg/ml). This  casts some doubts on the effectiveness of the employed all-atom force-field  to properly model coaxial stacking.

Finally, our work draws attention to the errors  affecting the estimate of the average chain length
$M$ via a straightforward comparison of the nematic coexisting concentrations with
analytic predictions based on the original Onsager theory for mono-disperse thin rods~\cite{BelliniScience07,MezzengaLangmuir2010}. 
We have found that such approximation significantly underestimates $M$ at the I-N transition concentration $c_N$. 
In addition, the theoretical approach predicts  a  re-entrant behavior of the transition lines in the $c$-$M X_0$ plane, a distinct feature of the polydisperse nature of the equilibrium chains.  

\section*{ACKNOWLEDGMENTS}
We thank Thomas Ouldridge, Flavio Romano  and Teun Vissers for fruitful discussions. We acknowledge support from ERC-226207-PATCHYCOLLOIDS and ITN-234810-COMPLOIDS as well as from NVIDIA.

\appendix
\section{Excluded volume contributions}
\label{sec:appendix_a}
Here we further discuss the calculation of the excluded volume term $v_{excl}(l,l')$ for the present model.
Following Ref.~\cite{ourMacromol}, the excluded volume is assumed to be the following
second order polynomial in $l$ and $l'$:
\begin{eqnarray}
&&v_{excl}[l,l'; f(\theta)] = 2  \int  f(\theta) f(\theta') D^3 \left [ \Psi_1(\gamma,X_0) + \right . \nonumber \\ 
&&+\frac{l + l'}{2} \Psi_2(\gamma,X_0) X_0 +\left. \Psi_3(\gamma,X_0) 
X_0^2\; l\, l' \right ] \nonumber\\
&&d{\boldsymbol\Omega}\, d{\boldsymbol\Omega}'
\label{eq:genvexcl}
\end{eqnarray}
where % $f(\theta)$ is the probability for a given monomer of having an orientation ${\bf u}$ within the solid angle ${\Omega}$ and ${\Omega}+d{\Omega}$ and
 the functions $\Psi_\alpha$, $\alpha=1,2,3$, describe the angular dependence of the excluded volume.
The orientational probability $f(\theta)$ is normalized such that
\begin{eqnarray}
\int f(\theta) d\boldsymbol\Omega = 1.
\label{eq:normcond}
\end{eqnarray}

The three contributions to the excluded volume in Eq.~(\ref{eq:genvexcl}) come from end-end, end-midsection and
midsection-midsection steric interactions~\cite{ourMacromol} between two chains. 
%and this expression, though exact only for two cylinders, is justifiable even for semi-flexible chains of HCs.

In the isotropic phase the orientational distribution does not have any angular dependence, i.e. $f(\theta)=1/4\pi$, and Eq. (\ref{eq:genvexcl})
reduces to the form
\begin{eqnarray}
%v_{excl}(l,l') &=& \frac{\pi^2}{8} D^3 + \left (\frac{3\pi}{8} + \frac{\pi^2}{8}\right) (l+l') X_0 D^3 +\nonumber\\ 
%&+& \frac{\pi}{2} l\,l' X_0^2 D^3 
v_{excl}(l,l',X_0) &=& B_I X_0^2 \,l\, l' +k_I v_d \frac{l+l'}{2} + A_I.
\label{eq:vexclisoA}
\end{eqnarray}

\begin{figure}[tbh]
\vskip 1cm
\includegraphics[width=.495\textwidth]{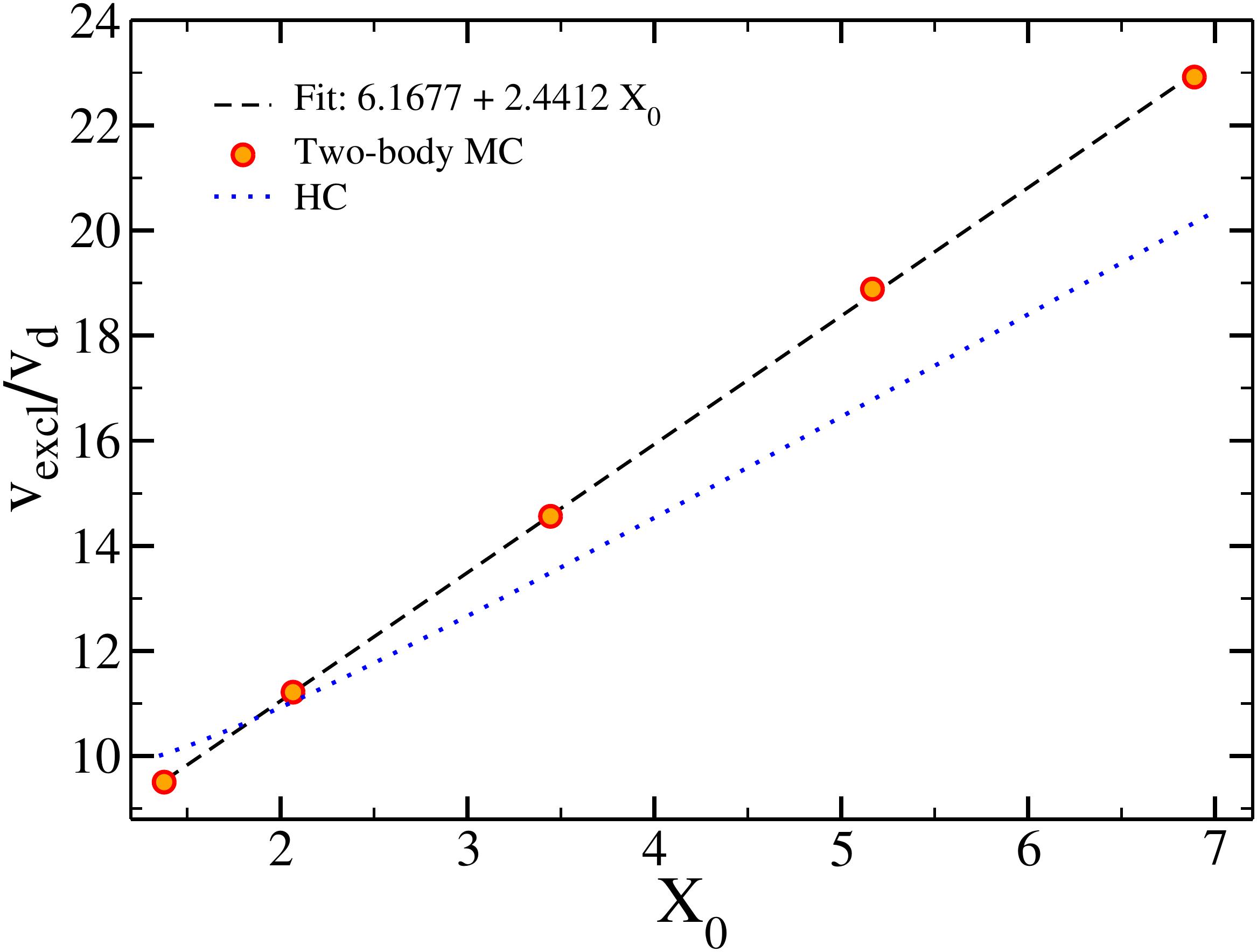}
\caption{Excluded volume in the isotropic phase together with analytic approximations. From the linear
fit one has $B_I= 0.959 D^3$ and $k_I=3.084$, while we assume $A_I=0$}
\label{fig:vexcliso}
\end{figure}

The parameters $B_I$, $k_I$ and $A_I$ appearing in Eq. (\ref{eq:vexclisoA}) can be calculated 
via MC integration procedures as discussed in the subsection~\ref{sec:twobody} and in Ref.~\cite{ourMacromol}.  We expect that
these parameters do not depend on $X_0$ because
each DNADs comprises $N_b$ stacked base pairs which are all identical with respect to excluded volume interactions
(i.e. they all have the same shape).
In particular, the calculated excluded volume of two DNADs is
reported in Fig.~\ref{fig:vexcliso} for $5$ different aspect ratios, together with the resulting values for the above parameters.   

Using the Onsager angular distribution $f_\alpha(\theta)$ in Eq.~(\ref{eq:vexclisoA}), the excluded volume in the nematic phase depends also on the parameter $\alpha$, i.e. the general form in Eq.~(\ref{eq:genvexcl}) reduces to
\begin{eqnarray}
%v_{excl}(l,l') &=& \frac{\pi^2}{8} D^3 + \left (\frac{3\pi}{8} + \frac{\pi^2}{8}\right) (l+l') X_0 D^3 +\nonumber\\ 
%&+& \frac{\pi}{2} l\,l' X_0^2 D^3 
v_{excl}(l,l',X_0,\alpha) &=& B_N(\alpha) X_0^2 \,l\, l' +k_N(\alpha) v_d \frac{l+l'}{2} \nonumber\\
 &+& A_N(\alpha).
\label{eq:vexclnemA}
\end{eqnarray}

\begin{figure}[tbh]
\vskip 1cm
\includegraphics[width=.495\textwidth]{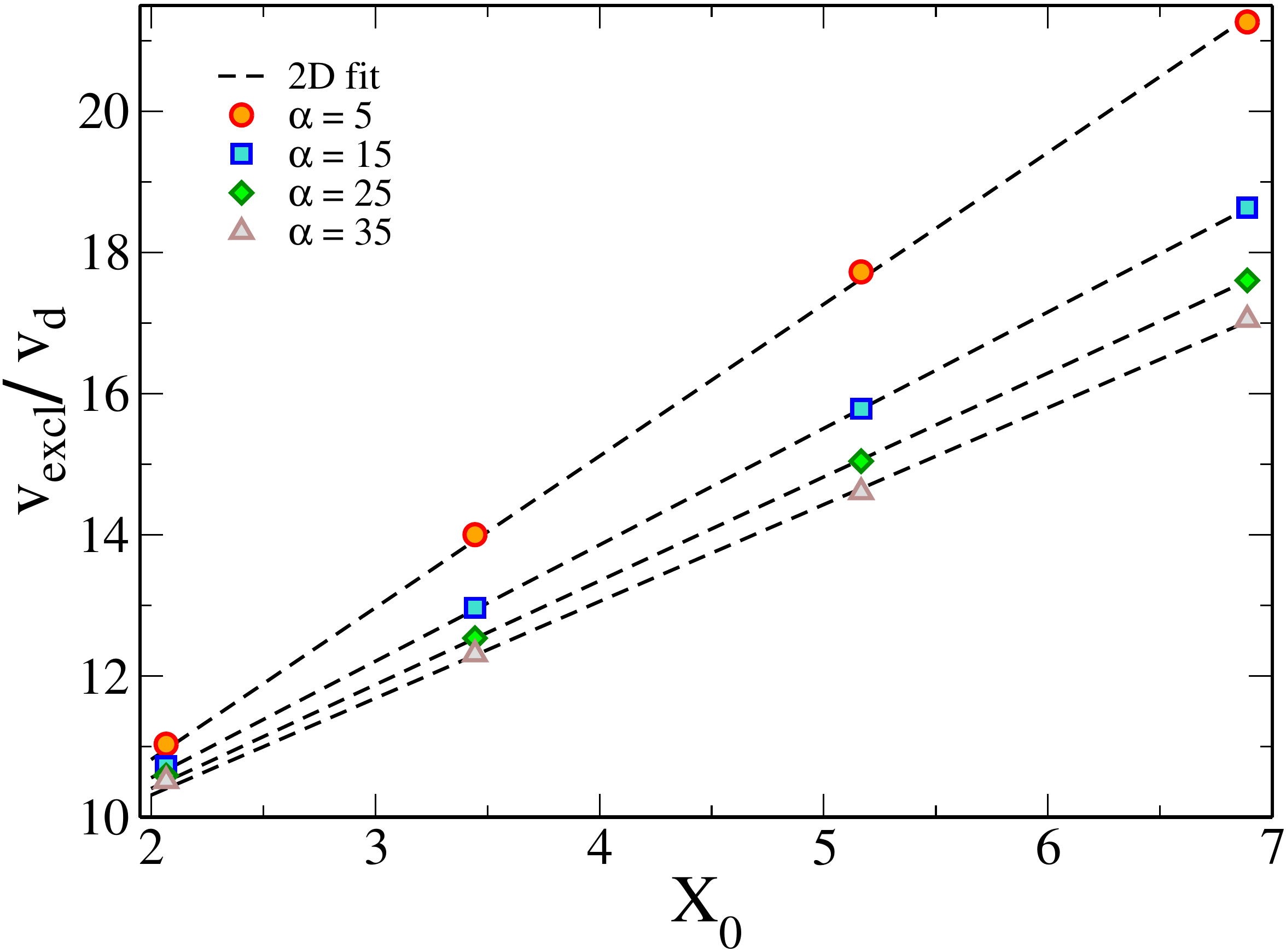}
\caption{Excluded volume as a function of aspect ratio $X_0$ in the nematic phase together with analytic approximations for several $\alpha$. The dashed lines are obtained plotting the function reported in Eq.  (\ref{eq:vexclnem}) and setting $\eta_1=0.386419$, 
$\eta_2=1.91328$ and $\eta_3=-0.836354$.}
	\label{fig:vexclnem}
\end{figure}

Assuming that $A_N(\alpha)=0$, $k_N(\alpha)=k_N^{HC}(\alpha)$ and $B_N(\alpha)$ is given by
%<<<<<<< .mine
%Eq. (\ref{eq:BN}), the three parameters $\eta_k $ with $k=1,2,3$ have to be estimated.
%For $l=l'=1$ and several values of $\alpha$ ($\alpha=5\dots 45$ in steps of $5$) and $X_0$ we calculated numerically the nematic excluded volume for two DNADs as shown in Fig.~\ref{fig:vexclnem}, where we plot  $v_{excl}/v_d$ vs $X_0$ for various $\alpha$.
%The dashed lines shown in Fig.~\ref{fig:vexclnem} are obtained through a two-dimensional fit 
%of our numerical data for $v_{excl}(1,1,X_0,\alpha)$ using Eq. (\ref{eq:vexclnem}) as the fitting function.
Eq. (\ref{eq:BN}), the three parameters $\eta_k $ with $k=1, 2, 3$ have to be estimated.
For $l=l'=1$ and several values of $\alpha$ ($\alpha=5\dots 45$ in steps of $5$) and $X_0$ we calculated numerically the nematic excluded volume for two DNADs. The results are shown in Fig.~\ref{fig:vexclnem}, where we plot  $v_{excl}/v_d$ vs $X_0$ for various $\alpha$.
The dashed lines shown in Fig.~\ref{fig:vexclnem} are obtained through a two-dimensional fit to numerical data for $v_{excl}(1,1,X_0,\alpha)$ using Eq.~(\ref{eq:vexclnem}) as fitting function.

\section{Excluded volume of hard cylinders}
\label{sec:appendix_b}

For two rigid chains of length $l$ and $l'$ which are composed of hard cylinders (HCs) of diameter $D$ and length $X_0 D$,  $v_{excl}(l,l')$ can be described by 
\begin{eqnarray}
v_{excl}^{HC}[l,l'; f(\theta)] &=& \int f(\theta) f(\theta') D^3 \left [ \, \frac{\pi}{2} \sin\gamma + \frac{\pi}{2} X_0\right . \nonumber\\
  && ( 1 + |\cos\gamma|  +  \frac{4}{\pi} E(\sin\gamma) ) + \nonumber\\
&+& \left .  \frac{l  + l'}{2} + 2 X_0^2 \sin \gamma\;\; l\,l'\, \right ] d{\boldsymbol\Omega}\, d{\boldsymbol\Omega}'
\label{eq:vexclHC}
\end{eqnarray}
where $\cos\gamma={\bf u}\cdot{\bf u}'$, $\bf u$ and $\bf u'$ are the orientations of two HCs and
$E(\sin\gamma)$ is the complete elliptical integral
\begin{equation}
E(\sin\gamma) = \frac{1}{4} \int_0^{2\pi} (1-\sin^2\gamma \sin^2\psi)^{1/2} d\psi.
\label{eq:ellipint}
\end{equation}

The integrals in Eq. (\ref{eq:vexclHC}) can be calculated exactly in the isotropic phase, 
while in the nematic phase the calculation can be done analytically only for suitable choices
of the angular distribution $f(\theta)$.  Here we assume that the angular distribution 
is given by the Onsager function in Eq.~(\ref{eq:fons}).
%Comparing Eqs. (\ref{eq:vexclHC})  and (\ref{eq:genvexcl}) for HC one has:
%\begin{eqnarray}
%\Psi_1(\gamma,X_0) &=&  \frac{\pi}{4} \sin\gamma \nonumber\\
%\Psi_2(\gamma,X_0) &=& \frac{\pi}{4} (1 + |\cos\gamma|  + \frac{4}{\pi} E(\sin\gamma) ) \nonumber\\
%\Psi_3(\gamma,X_0) &=& \sin\gamma
%\label{eq:psiforHC}
%\end{eqnarray}
%In view of Eqs. (\ref{eq:psiforHC}) we note that for HCs the functions $\Psi_1(\gamma)$, $\Psi_2(\gamma)$ and $\Psi_3(\gamma)$
%accounts for the orientational dependence of the excluded volume of two monomers having orientations ${\bf u}$ and ${\bf u}'$
%with ${\bf u}\cdot{\bf u}' =\cos\gamma$.

Using the Onsager orientational function the following approximate expressions 
for the coefficients $k_N(\alpha)$, $B_N(\alpha)$ and $A_N(\alpha)$ can be derived~\cite{Odijk86}
\begin{eqnarray}
\tilde B_N(\alpha) &=&  D^3 (\pi/4)\rho_a(\alpha) \nonumber\\
\tilde k_N(\alpha) &=& \pi D^3\frac{X_0}{v_d}\left(1-\frac{1}{\alpha }\right)   \nonumber\\
\tilde A_N(\alpha) &=& D^3\left(\pi/4\right )^2\rho_a(\alpha)
\label{eq:approxparams}
\end{eqnarray}
where 
\begin{eqnarray}
\rho_a&=& 4 (\pi \alpha )^{-1/2}\left(1 - \frac{15}{16\,\alpha  }+\frac{105}{512\,\alpha ^2}\right .+\nonumber\\
&& \left .+\frac{315}{8192\,\alpha ^3}\right)
\end{eqnarray}

We evaluate numerically %over a given grid of points using 
%the Onsager orientational function defined in Eq. (\ref{eq:fons}), and then we fit the numerical data using the following expressions for $A_N$, $B_N$ and $f_N$:
the excluded volume in Eq. (\ref{eq:vexclHC}) for many values of $\alpha$ and,  building on the expressions in Eqs. (\ref{eq:approxparams}),
 we perform a fit to 
this data using the following functions:
\begin{eqnarray}
B_N^{HC} (\alpha)&\simeq& D^3 (\pi/4) \left ( \rho_a(\alpha) + \frac{c_4}{\alpha ^{9/2}}+\frac{c_5}{\alpha ^{11/2}}\right )
\\ 
k_N^{HC}(\alpha) &=& 4 \left(1-\frac{1}{\alpha }\right)+ \sum_{i=2}^{\infty}\frac{b_i}{\alpha ^i}\simeq \frac{4}{\pi} \sum_{i=0}^4 \frac{d_i}{\alpha^i}\\
A_N^{HC}(\alpha) &\simeq& D^3 (\pi/4)^2 \left ( \rho_a(\alpha) + \frac{c_4}{\alpha ^{9/2}}+\frac{c_5}{\alpha ^{11/2}}\right )
\label{eq:nemparamsfits}
\end{eqnarray} 
%<<<<<<< .mine
%{\bf where $c_m$ with $m=4,5$ and $d_i$ with $i=0\ldots 4$ are fit parameters. 
%The values for this parameters which are obtained from the fitting procedure are 
% $c_4=1.2563$, $c_5=-0.95535$, $d_0=3.0846$, $d_1=-4.0872$, $d_2=9.0137$, $d_3=-9.009$, $d_4=3.3461$.}
%=======

The coefficient values resulting from the fitting procedure are $c_4=1.2563$, $c_5=-0.95535$, $d_0=3.0846$, $d_1=-4.0872$, $d_2=9.0137$, $d_3=-9.009$ and $d_4=3.3461$.

%>>>>>>> .r343
%{PF2 -> 3.08456, PF3 -> -4.08721, PF4 -> 9.01371, PF5 -> -9.00902, 
% PF6 -> 3.34611}

\end{document}